\documentclass[a4paper,twocolumn,eqsecnum,eprintnumbers,floatfix,aps,prd,nofootinbib,amsmath,amssymb,amsfonts,superscriptaddress,twoside]{revtex4-2}
\usepackage{graphicx,rotating,slashed,amsmath,xcolor,colortbl}
\usepackage{tabularx}
\usepackage[utf8]{inputenc} 
\usepackage[T1]{fontenc}

\usepackage{comment}
\usepackage{bm}
\usepackage[normalem]{ulem}
\usepackage{tikz}
\usepackage{mciteplus} 
\usepackage{enumitem}
\usepackage{booktabs}
\usepackage{bigints}
\usepackage{float} 

\allowdisplaybreaks	 

\newcommand{\bbeta}{{\bm{\eta}}}
\newcommand{\bx}{{\bf{x}}}

\newcommand{\bb}{{\bf{b}}}

\def\ii{{\rm i}} 
\def\e{{\rm e}} 

\def\bea{\begin{eqnarray}}
\def\eea{\end{eqnarray}}

\def\dd{{\rm d}}

\def\nn{\nonumber}

\newcommand{\be}{\begin{equation}}
\newcommand{\ee}{\end{equation}}

\def \bm#1{\mbox{\boldmath$#1$\unboldmath}} 

\definecolor{rossos}{cmyk}{0,1,1,0.55}
\definecolor{blu}{cmyk}{1,1,0,0.3}
\definecolor{bluc}{cmyk}{1,1,0,0.1}
\definecolor{verde}{cmyk}{0.92,0,0.59,0.25}
\definecolor{verdec}{cmyk}{0.92,0,0.59,0.15}
\definecolor{verdes}{cmyk}{0.92,0,0.59,0.4}
\usepackage{hyperref}
\hypersetup{colorlinks, bookmarksopen, bookmarksnumbered,
citecolor=blu, linkcolor=bluc, pdfstartview=FitH, urlcolor=rossos}

\definecolor{mygreen}{rgb}{0,0.6,0}

\newcommand{\cpi}[1]{{\color{blue} #1}}

\begin{document}
\title{Beyond-eikonal diffraction integral in gravitational lensing}
\author{Emma Bruyere}
\email{bruyere@iap.fr}
\author{Giulia Cusin}
\email{cusin@iap.fr}
\affiliation{Institut d'Astrophysique de Paris, UMR-7095 du CNRS et de Sorbonne Universit\'e, Paris, France}
\affiliation{Département de Physique Théorique and Center for Astroparticle Physics, Université de Genève, Quai E. Ansermet 24, CH-1211 Genève 4, Switzerland}
\author{Cyril Pitrou}
\email{pitrou@iap.fr}
\affiliation{Institut d'Astrophysique de Paris, UMR-7095 du CNRS et de Sorbonne Universit\'e, Paris, France}

\begin{abstract}
\noindent 
We revisit the derivation of the diffraction integral, which is an approximate evaluation of the Kirchhoff integral, widely used in the literature for phenomenological applications in gravitational lensing. We propose a systematic approach to evaluate the Kirchhoff integral within a beyond-eikonal expansion, carefully tracking all approximations involved in its standard derivation. In this framework, we recover the usual diffraction integral as the leading contribution, together with a correction term that is typically small in realistic lensing scenarios. We further show that our evaluation of the Kirchhoff integral is equivalent to a beyond-eikonal expansion of wave propagation, after summing over all geometric-optics images. This result clarifies the physical origin of diffraction in gravitational lensing and demonstrates that diffraction effects and beyond-eikonal corrections are not distinct phenomena, but rather different but equivalent descriptions of the same underlying physics.
\end{abstract}


\maketitle

\section{Introduction}

Gravitational lensing~\cite{SEF1992,Bartelmann:2010fz,Narayan:1996ba} is a well-established phenomenon in light propagation and is expected to affect gravitational waves (GWs) as well. Accordingly, its impact is being actively investigated using current and forthcoming GW observations 
\cite{Sereno:2010dr,Jung:2017flg,Lai:2018rto,Oguri:2018muv,Dai:2018enj,Diego:2019lcd,Hannuksela:2019kle,Liu:2020par,Urrutia:2021qak,LIGOScientific:2021izm,LIGOScientific:2023bwz,Yin:2023kzr, Cusin:2020ezb, Pijnenburg:2024btj, Cusin:2021rjp, Cusin:2019eyv, Sberna:2022qbn,Toubiana:2020drf,Toscani:2023sfc}. Since the wavelength of GWs can be comparable to the characteristic size of astrophysical lenses, the geometric-optics (GO) approximation may break down in a variety of physically relevant scenarios. As a result, wave diffraction plays a central role in the gravitational lensing of GWs. This regime is typically studied within the scalar approximation, in which the tensor-valued GW field is replaced by a scalar field, thereby neglecting polarization effects.
The Kirchhoff integral is a reformulation of scalar wave propagation in a stationary spacetime that allows us to interpret wave propagation in terms of diffraction effects.  In this article, we revisit the derivation of the lensing amplification factor of a scalar field using the Kirchhoff integral. 

The usual evaluation of the Kirchhoff integral in this context relies on two key assumptions. First, the lens--source and lens--observer distances are much larger than the typical transverse size of the beam. This implies that all propagating directions from the source to the lens, or from the lens to the observer, can be assumed to be parallel. Barring phase effects, all points in the lens plane can be taken as equidistant from the source and observer; thus, the wave amplitude is assumed to be uniform across the lens plane. The second assumption is that the wave, when crossing the lens plane, is modified only by the GO time delay (also called Shapiro time delay~\cite{1964PhRvL..13..789S}), that is by a frequency-independent phase. 

Under these approximations, the Kirchhoff integral reduces to an integral of the exponential of the time-delay function over the lens plane. We refer to this quantity as the \emph{diffraction integral}~\cite{Bliokh1975,Bontz:1981rvr,1985A&A...148..369S,Deguchi:1986zz,1986ApJ...307...30D}. Despite the number of assumptions involved, the diffraction integral is typically assumed to be valid for all values of $\omega$, where $\omega$ is the angular frequency of the signal, ranging from the geometric-optics limit to the deep wave-optics regime. A vast literature has been dedicated to lensing phenomenology using this expression, see e.g., Refs.~\cite{Takahashi:2003ix, Takahashi:2004mc, Feldbrugge:2020ycp, Cremonese:2021puh,Caliskan:2022hbu,Jow:2022pux, Savastano:2023spl,Yeung:2024pir,Villarrubia-Rojo:2024xcj,Zumalacarregui:2024ocb, Cheung:2024ugg, Zumalacarregui:2026uqs, Caliskan:2023zqm, Chen:2024xal, Ezquiaga:2025gkd,Tambalo:2022plm, Tambalo:2022wlm,Braga:2024pik,Bonga:2024orc,Vujeva:2025nwg,Vujeva:2025kko,Ezquiaga:2020gdt}. 

We propose a new method to evaluate the Kirchhoff integral based on the beyond-eikonal (BE) expansion. It allows to keep track of amplitude variations across the lens plane, and it avoids assuming that all propagating directions are parallel. Our result captures diffraction effects beyond the simple GO time delay, and contains the usual diffraction integral as its leading contribution~\cite{Takahashi:2004mc}, together with an additional correction term. This correction is relevant near the lens plane but fades away with increasing distance from it. We show that it is typically very small in realistic lensing configurations, implying that the diffraction integral provides an excellent approximation within the regime of large frequencies (i.e., the BE regime) considered here. Explicitly, using standard notation of lensing amplification (see the text for definitions), we find for images that did not cross a caustic that the amplification factor receives a correction of order $\mathcal{O}\left(\omega^{-1}\right)$ of the form 
\begin{align}\label{ImprovedF}
F(\omega, \bbeta)&=\sum_j \sqrt{\mu_j} \left(1+\frac{\ii}{\omega}\Delta^{\text{st}}_j+\frac{\ii}{\omega}\Delta^{\text{new}}_j\right) e^{\ii \omega T_j} \,,
\end{align}
where the sum is over the GO images (extrema of the time delay function) and $\omega$ is the angular frequency, which must satisfy $\omega_{\rm ref}/\omega \ll 1$. Here $\omega_{\rm ref}$ is the frequency at which the Fresnel zone size is comparable to the characteristic lens scale. This frequency separates the regime of deep wave optics from the regime of BE propagation. The leading order term in $1/\omega$ corresponds to the GO result which accounts for a lensing magnification of each image, and a GO (frequency-independent) time delay. The other two terms are the leading order corrections appearing in the BE expansion. Both $\Delta^{\text{st}}_j$ and $\Delta^{\text{new}}_j$ are real quantities, so these corrections give a frequency-dependent correction to the GO time delay. The first correction is the standard one arising in the diffraction integral, while the second correction is a new term, which we derive in this article. These contributions at linear order in $G$ have the form
\begin{align}
&\Delta^{\text{st}}_j=\pi G \nabla^2_\perp\Sigma \left( \frac{D_{\rm L} D_{\rm LS}}{D_{\rm S}}\right)^2 \,,\\
&\Delta^{\text{new}}_j=2\pi G \Sigma  \frac{D_{\rm LS}^2}{D_{\rm S}^2}\,,
\end{align}
where $\Sigma$ is the projected surface density and $\nabla_\perp$ is the flat derivative in the two-dimensional lens plane. Here, $D_{\rm L}$, $D_{\rm LS}$ and $D_{\rm S}$ are the observer-lens, lens-source, and observer-source distances, respectively, measured along the geometric-optics trajectory.\footnote{For small impact parameter, in first approximation these can be identified with the distance between the observer and the lens plane, the distance between the lens and source planes, and the distance from the observer to the source plane.} The new term $\Delta^{\text{new}}_j$ arises from amplitude variation across the lens plane and projection effects usually neglected in the standard Kirchhoff evaluation. It is typically suppressed by a factor of $(b/D_{\rm L})^2$ with respect to the standard term, where $b$ is the characteristic scale of the lens gravitational potential. Therefore, unless the lens is very close to us, this additional term can be neglected in realistic lensing phenomenological applications. 

We also show (at linear order in the lens potential $\phi$) that propagating the field from the source to the observer using the perturbative BE approach of \cite{Bruyere:2026gnt}, and summing over all geometric-optics images, is equivalent to our improved evaluation of the Kirchhoff integral. In the BE approach of \cite{Bruyere:2026gnt}, the neighborhood of each geometric-optics trajectory is described by a Taylor expansion of all relevant quantities around a given GO geodesic.
In the Kirchhoff formalism, by contrast, we integrate over an arbitrary surface close to the lens plane, and contributions arising from regions around the GO image positions dominate the integral. We show that the physics captured by both methods is exactly the same\footnote{The equivalence is exact when working at linear order in $G$, i.e., in the absence of caustics. When caustics are present, the BE expansion is not able to unambiguously fix the phase, contrary to the Kirchhoff integral formalism.} provided that the former is used to set the conditions of the field on the surface on which the Kirchhoff integral is performed. This is a nontrivial result that helps clarify the physical origin of diffraction effects.

This equivalence has not been clearly emphasized in the literature and has sometimes led to confusion. Diffraction and BE effects are often discussed as if they were distinct phenomena. Our results show that this is not the case. Upon entering the BE regime, that is when accounting for corrections to the GO result, diffraction naturally appears, and the effects captured by the Kirchhoff formalism are precisely those obtained by the BE expansion.

The paper is structured as follows. In section \ref{OldNew} we review the standard derivation of the diffraction integral from the Kirchhoff integral and we discuss its BE limit. In section \ref{previous} we review the results of \cite{Bruyere:2026gnt}, where the equation of motion for a scalar field is solved perturbatively in the BE expansion. In section \ref{improved} we derive the formal expression of the improved Kirchhoff integral, and in section~\ref{SecGauss} we evaluate it at linear order in the gravitational potential for a lens with non-vanishing matter density. We also exhibit how the diffraction integral approximation is contained in our improved evaluation, hence showing that there are new BE terms not accounted for in the standard treatment. In section \ref{SIS} we present an application to a realistic lens model, and in section \ref{discussion} we discuss our results. Intermediate technical results are gathered in a series of appendices.

\section{Old concepts and new ideas}\label{OldNew}

\subsection{Standard Kirchhoff integral}

The propagation of a scalar wave is dictated by the Klein-Gordon equation
\be\label{EqCurvedBox}
\Box \hat{\Psi}=0\,,
\ee
where $\Box$ is the d'Alembertian in a spacetime with the gravitational potential of the lens. For a monochromatic wave in a stationary spacetime, we separate the spatial and temporal dependences according to 
\be\label{TotPsivsPsi}
\hat{\Psi}=e^{\cpi{-}\ii\omega t} \Psi\,,
\ee
where $\Psi$ depends only on space. 
Far from the lens plane, the gravitational potential can be neglected and the field satisfies  
\be\label{EqPsi}
(\Delta + \omega^2)\Psi=0 \,,
\ee
where $\Delta$ is the flat space Laplacian.

\begin{figure}[!ht]
\includegraphics[width=0.52\columnwidth]{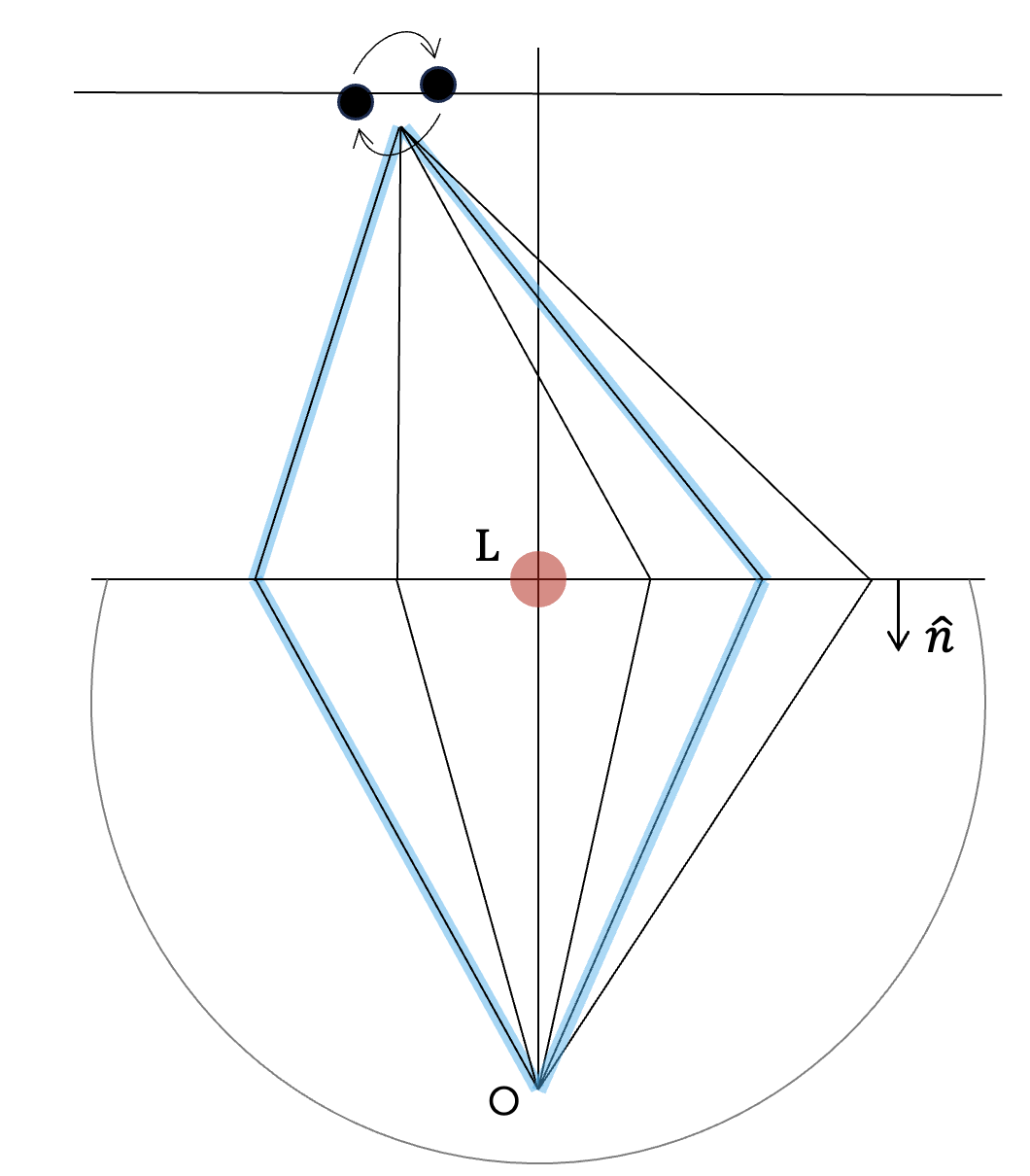}
\includegraphics[width=0.46\columnwidth]{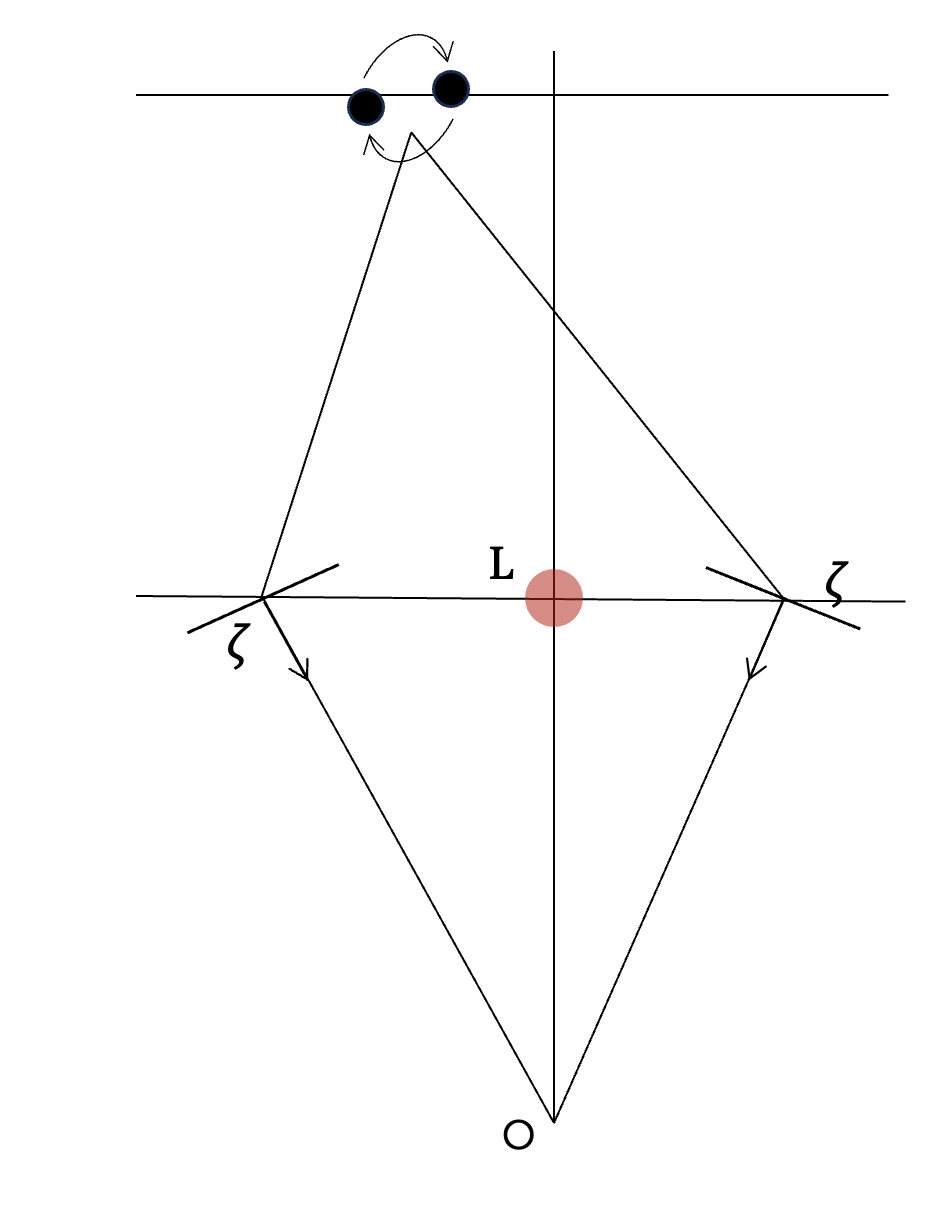}
\caption{Kirchhoff integral approximations.
 {\it Left}: Standard diffraction integral approximation. We choose a surface enclosed by the (offset) lens plane and a sphere of radius $R\rightarrow \infty$ that contains the observer. The field at the observer is then obtained as a surface integral of the field on the (offset) lens plane. This neglects projection effects and only considers distance differences in the phase. The field at the observer is therefore written as an integral over the exponential of the time delay associated with all possible paths intersecting the lens plane. When evaluated BE, this approach describes diffraction around the geometric-optics trajectories (shown in blue).
{\it Right}: In our improved evaluation, the Kirchhoff integral is evaluated only locally around the extrema trajectories. The field in the (local) integration plane is obtained by evolving the BE equations from the source, and projection effects are consistently taken into account.}
\label{Lens0}
\end{figure}

The Kirchhoff integral, derived from this, gives the field at the observer as an integral on an arbitrary closed surface around a flat region containing the observer (see Appendix \ref{Kirchhoff} for a derivation). In full generality the Kirchhoff integral is expressed as
\be\label{surfText}
\Psi = \frac{1}{4 \pi}\int_{S} \left[\Psi \partial_n \left(\frac{{\rm e}^{\ii \omega \ell}}{\ell}\right)-\left(\frac{{\rm e}^{\ii \omega \ell}}{\ell}\right)\partial_n \Psi\right] \dd^2 S \,,
\ee
where $\bm n$ is the inward-pointing unit normal, $\partial_n$ is the directional derivative along this unit vector, and $\ell$ is the distance from the observer to a point on $S$.

This expression is general and independent of the explicit form of $S$; however, it relies on the assumptions that the enclosed region is flat. To evaluate (\ref{surfText}), one must explicitly choose the most convenient form of the surface and specify the field $\Psi$ and its derivative $\partial_n \Psi$ on it.

\subsection{The diffraction integral approximation}

 A common choice is to consider a plane $E'$ orthogonal to the optical axis (and parallel to the lens plane $E$), sufficiently far away so that the spacetime on $E'$ can be treated as flat, but still sufficiently close so that the distances from the observer to $E$ and to $E'$ can be identified. Then $S$ is chosen to be the surface enclosed by $E'$  and a sphere of radius $R$ centered on the observer, see Fig.~\ref{Lens0}. One then takes the limit $R \to \infty$ in such a way that the only part of the surface $S$ on which the field $\Psi$ does not vanish is the plane $E'$. In this limit, the integral reduces to a planar one.
 Choosing a system of coordinates with the polar axis aligned with the optical axis (see left panel of Fig.~\ref{Lens0}), Eq.~(\ref{surfText}) reduces to 
 \be\label{Kirchhhoff1}
\Psi = \frac{1}{4 \pi}\int_{E'} \left[\Psi \partial_z \left(\frac{{\rm e}^{\ii \omega \ell}}{\ell}\right)-\left(\frac{{\rm e}^{\ii \omega \ell}}{\ell}\right)\partial_z \Psi\right] \dd^2 \bm{b} \,,
\ee
where $\bm{b}$ is the two-dimensional impact vector on the lens plane. 
One must then specify the form of the field on $E'$, and this step requires approximations. Traditionally, we assume that the propagation from the source to $E'$ can be described within the geometric-optics approximation. The field is therefore parametrized by a slowly varying amplitude $A$ and a fast evolving phase as
\be
\Psi=A e^{\ii \omega\varphi}\,.
\ee
With~\eqref{TotPsivsPsi} the total phase $S$ is defined as 
\be\label{phase_S}
\omega S\equiv \omega(\varphi - t)\,,
\ee
and its gradient $k_a \equiv \partial_a S$ is a null vector ($k^a k_a=0$) that defines null geodesics ($k^a \nabla_a k^b = 0$)~\cite{Isaacson:1967zz,Cusin:2024git}. Equivalently, the phase is constant along these null geodesics since $\dd S/\dd s = k^a \partial_a S = k^a k_a =0$, where $s$ is the affine parameter.

\begin{table*}[ht!!]
\centering
\begin{tabular}{l|l}
\textbf{symbol} & \textbf{definition} \\
\hline
$\ell$& distance from the observer to any point \\
$d$& distance from the source to any point \\
$D_{\rm L}$& distance from the observer to a generic point $L$ on the lens plane\\
$\bar{D}_{\rm L}$& distance from the observer to the lens plane\\
$D_{\rm L}^{\text{str}}$ & distance to the lens plane via the straight trajectory of the flat space geodesic\\
$D_{\rm L}^{\text{curv}}=D_{\rm L}-D_{\rm L}^{\text{str}}$ & difference of the distance to the lens plane with respect to the flat space reference\\
$D_{\rm LS}$& distance from the source to a generic point on lens plane\\
$\bar{D}_{\rm LS}$& distance from the source plane to the lens plane\\
$D_{\rm LS}^{\text{str}}$ & distance from the source to the lens plane via the straight trajectory of the flat space geodesic\\
$D_{\rm LS}^{\text{curv}}=D_{\rm LS}-D_{\rm LS}^{\text{str}}$ & difference of the distance from the source to the lens plane with respect to the flat space reference\\
$D_K$ & distance from the observer to a point $K$ located on the geodesic after the lens\\
\hline
\end{tabular}
\caption{Summary of the various distance definitions.}
\label{tab:mytable}
\end{table*}

\begin{figure}[!ht]
\centering
\includegraphics[width=0.75\columnwidth]{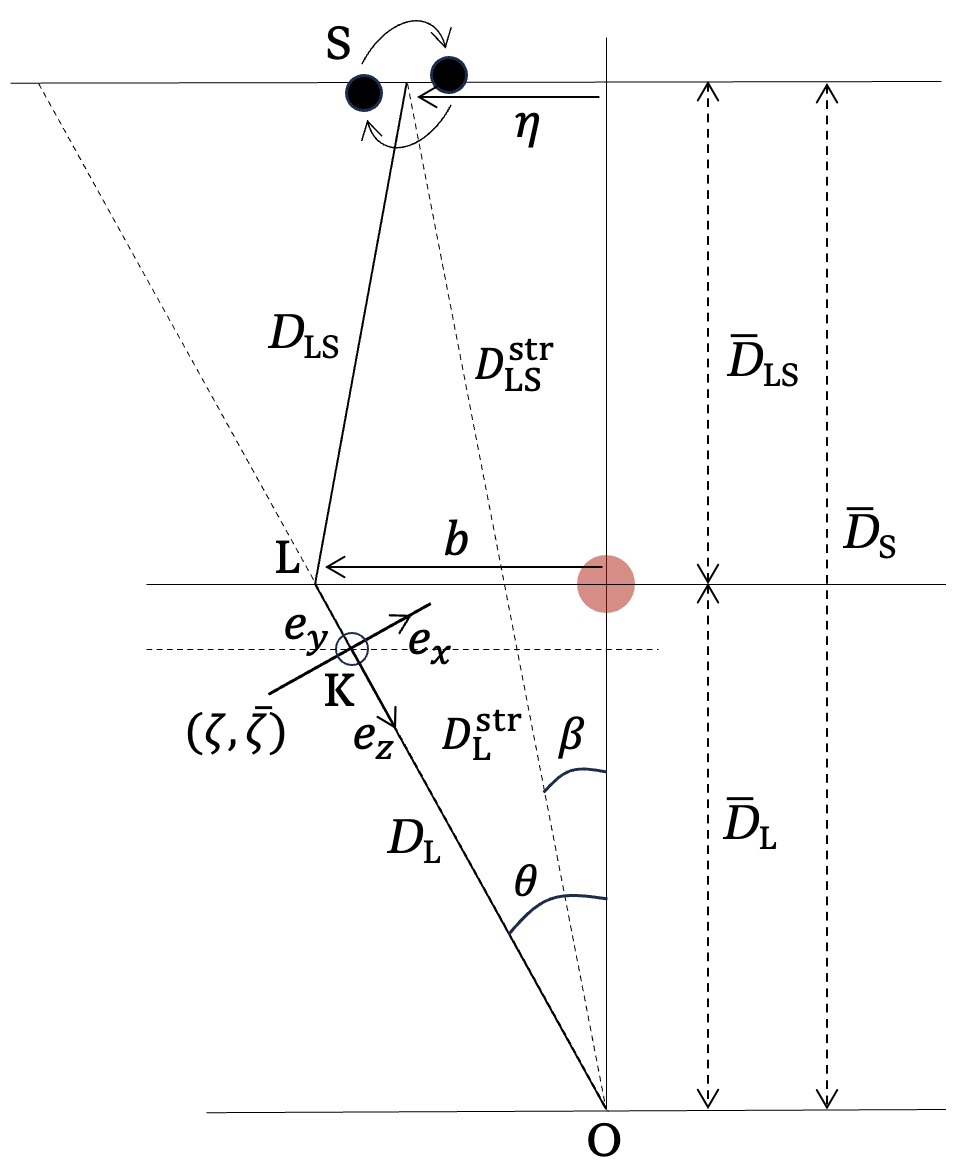}
\caption{\label{setup}Schematic representation of the system geometry, with the various distances and the coordinates chosen when performing the integral around the GO trajectory, see section \ref{improved}.}
\end{figure}

Let us focus on the form of the field on the lens plane obtained under these assumptions. In the thin lens approximation, the space between the source and the lens plane is flat, and the phase emanating from a point source in flat space is of the form 
\be\label{PhaseFlat}
\omega S = \omega(d- t)\,,
\ee
where $d$ is the distance from the source, and the field amplitude of the spherical wave scales as $A \propto 1/d$. In the curved spacetime (with the lens), the geodesics differ from those in flat spacetime, which are straight lines, hence surfaces of constant phase differ. We introduce a source plane parallel to the lens plane $E$, and two-dimensional vectors $\boldsymbol{b}$ and $\bm{\eta}$ representing the impact parameter in the lens plane and the source position in the source plane, respectively. We denote by $\bar D_{\rm L}$, $\bar D_{\rm S}$ and $\bar D_{\rm LS}$ the distances between observer and lens plane, observer and source plane, and lens plane and source plane, respectively.  A summary of these distances is given in table\,\ref{tab:mytable} and Fig.\,\ref{setup}. We further decompose the distances into a flat-spacetime contribution (associated with a straight line connecting the source to the observer) and an additive correction which appears in curved space due to lensing deviation. Explicitly, this decomposition reads $D_{\rm LS}=D_{\rm LS}^{\rm str}+D_{\rm LS}^{\rm curv}$  and $D_{\rm{L}}=D_{\rm L}^{\rm str}+D_{\rm L}^{\rm curv}$, where "str" is the flat space (straight line) reference and "curv" is the correction in curved space. These distances are represented in Fig.~\ref{setup} and satisfy 
\begin{align}
D_{\rm LS} &= \sqrt{\bar{D}_{\rm LS}^2 + ({\bm b}-{\bm \eta})^2}\,,\\
D_{\rm{L}}&=\sqrt{\bar D_{\rm L}^2+\bm b^2}\,,\\
D^{\rm str}_{\rm LS} &= \bar{D}_{\rm LS}\sqrt{1 + (\bm \eta/\bar{D}_{\rm S})^2}\,,\\
D_{\rm L}^{\rm str} &= \bar D_{\rm L} \sqrt{1+(\bm \eta/\bar{D}_{\rm S})^2}\,.
\end{align}
Second, we assume that the effect of the lens can be approximated by an additional gravitational time delay, whose expression to linear order in the gravitational potential $\phi$ is 
\be\label{TGrav}
T_{\rm grav} = -2 \int \phi\, \dd s \equiv - \psi\,.
\ee

With the previous notation, the total time delay with respect to the flat spacetime trajectory is
\begin{equation}\label{timed}
T(\bm{b}, \bm{\eta}) =T_{\text{geom}}(\bm{b}, \bm{\eta}) + T_{\text{grav}}(\bm{b})\,,
\end{equation}
where $T_{\text{grav}}$ is given by (\ref{TGrav}) while 
\begin{align}\label{Tgeom}
&T_{\text{geom}}(\bm{b}, \bm{\eta})=D_{\rm{L}}^{\rm curv} + D_{\rm{LS}}^{\rm curv}  \simeq \frac{\bar D_{\rm L} \bar D_{\rm S}}{2 \bar D_{\rm LS}}\left(\frac{\bm{b}}{\bar D_{\rm L}}-\frac{\bm{\eta}}{\bar D_{\rm S}}\right)^2 .
\end{align}
Note that the last approximation is a second order expansion in $\bm{b}$ and $\bm{\eta}$, hence quartic corrections and beyond are discarded.

Finally, the $\partial_n$ derivatives in \eqref{surfText} can be approximated by
\begin{align}
\partial_n \Psi &\simeq \ii \omega (\bm{n}\cdot \bm{\nabla} d)  \Psi\,,\\
\partial_n \left(\frac{{\rm e}^{\ii\omega \ell}}{\ell}\right) &\simeq \ii \omega \,\left(\bm{n}\cdot \bm{\nabla} \ell\right) \left(\frac{{\rm e}^{\ii \omega\ell}}{\ell}\right)\,,
\end{align}
since variations of distances or amplitude are suppressed with respect to variations of the phase by a factor $1/\omega$.

Since the gravitational potential satisfies $\Delta \phi = 4\pi G \rho_{\rm mat}$, the integrated potential $\psi$, which is also the deflection potential, satisfies  $\nabla^2_\perp\psi=8\pi G \Sigma$ where $\Sigma$ is the surface mass density of the lens. 
We further use the approximations $\bm{n} \cdot \bm{\nabla} d \simeq 1$ and $\bm{n} \cdot \bm{\nabla} \ell\simeq -1$, we neglect the amplitude variations across the lens plane, and we approximate $D_{\rm LS}^{\rm str} \simeq \bar D_{\rm LS}$ and $D_{\rm L}^{\rm str} \simeq \bar D_{\rm L}$. By combining these simplifications, \eqref{surfText} reduces to the standard integral
 \be\label{finalSt}
\Psi^{\text{st}}(0)=\frac{\omega A|_{\rm L} {\rm e}^{\ii \omega D_{\rm S}^{\rm str}} }{2\pi \ii \bar D_{\rm L}} \int {\rm d}^2\bm{b}\, e^{\ii \omega T}\,,
\ee
where $A|_L$ denotes the field amplitude on the lens plane.
We refer to this approximate expression as the \emph{diffraction integral}. 

\subsection{Standard amplification factor}\label{StF}

The amplification factor is the ratio of the lensed to the unlensed field. 
Since the unlensed field is a spherical wave, $\Psi_{\rm NL}= (\left.D_{\rm LS}^{\rm str} A\right|_{\rm L})\e^{\ii\omega D^{\rm str}_{\rm S}}/D_{\rm S}^{\rm str} \simeq (\left. \bar{D}_{\rm LS} A\right|_{\rm L})\e^{\ii\omega D^{\rm str}_{\rm S}}/\bar{D}_{\rm S} $ where in the last step we approximated distances with distances between planes in the amplitude part of the wave (but not in the phase). 
From the diffraction integral (\ref{finalSt}), the amplification factor reads 
\be\label{final2}
F^{\text{st}}(\omega, \bm{\eta}) =\frac{\bar D_{\rm S} }{\bar D_{\rm L}\bar D_{\rm LS}} \frac{\omega }{2\pi \ii}\int d^2\bm{b}\,e^{\ii \omega T(\bb, \bbeta)}\,. 
\ee
Note that with the chosen normalization, $|F|=1$ in the no-lens limit. This amplification factor underlies a vast literature on lensing phenomenology \cite{Takahashi:2003ix, Takahashi:2004mc, Savastano:2023spl,Yeung:2024pir,Villarrubia-Rojo:2024xcj,Zumalacarregui:2024ocb, Cheung:2024ugg, Zumalacarregui:2026uqs, Caliskan:2023zqm, Chen:2024xal, Ezquiaga:2025gkd,Tambalo:2022plm, Tambalo:2022wlm,Braga:2024pik}. However, it was derived under several assumptions (geometric-optics until the thin lens, projection effects neglected), so care must be taken when extrapolating it to any lensing regime. 

In the GO limit ($\omega T \gg 1$) the stationary points of $T$ dominate the integral, and the image positions are determined by the lens equation, $\left.\nabla_{\perp} T\right|_{b_j}=0$. This is simply Fermat’s principle. To evaluate BE corrections, one typically performs a Taylor expansion of the time delay function around the extrema~\cite{Takahashi:2004mc,Jow:2020rcy}.
This is represented in the left panel of Fig.~\ref{Lens0}. The next-to-leading order terms follow from a fourth-order expansion:
\begin{align}\label{tdex}
&T(\boldsymbol{b}_j, \bbeta)=T_j+\frac{1}{2}\sum_{ac}\partial_a\partial_c T(\boldsymbol{b}_j, \bbeta) \bar{b}_a \bar{b}_c\nn\\
&+\frac{1}{6}\sum_{acd}\partial_a\partial_c\partial_d T(\boldsymbol{b}_j, \bbeta) \bar{b}_a\bar{b}_c\bar{b}_d\nn\\
&+\frac{1}{24}\sum_{acde}\partial_a\partial_c\partial_d \partial_e T(\boldsymbol{b}_j, \bbeta) \bar{b}_a\bar{b}_c\bar{b}_d\bar{b}_e+\dots
\end{align}
where $\partial_a \equiv \partial/\partial b^a$ and $\boldsymbol{\bar b}\equiv \boldsymbol{b}-\boldsymbol{b}_j$. 
The diffraction integral can be written as a Gaussian integral of the form\footnote{We are neglecting terms with odd powers of $\bm{b}$ which yield a vanishing contribution once integrated.}
 \be\label{FFF}
F(\omega, \bm{\eta}) \simeq\frac{\bar D_{\rm S} }{\bar D_{\rm L}\bar D_{\rm LS}} \frac{\omega }{2\pi \ii} \int d^2\bar{\bm{b}}\,e^{\ii \omega T_{\text{quad}}}\left[1+
\frac{\ii}{\omega}\mathcal{T}(\bar{\bm{b}}) \right]\,. 
\ee
where $T_{\text{quad}}$ yields the GO result and $\mathcal{T} $ is the BE phase correction. 
Explicitly, introducing
\be\label{def_L}
\bar{L}=\bar D_{\rm L}\bar D_{\rm LS}/\bar D_{\rm S}\quad \Leftrightarrow\quad \frac{1}{\bar{L}} = \frac{1}{\bar{D}_{\rm LS}} + \frac{1}{\bar{D}_{\rm L}}\,,
\ee
the Hessian matrix reads
\be\label{DefHab}
(H_j)_{ab}=\partial_a\partial_b T=\bar{L}^{-1}\delta_{ab}-\partial_a \partial_b \psi\,,
\ee
the quadratic expansion of the time delay reads
\be \label{Tquad}
T_{\rm quad}=T_j+\frac{1}{2}\boldsymbol{\bar b}^TH_j\boldsymbol{\bar b} \,,
\ee
and the BE correction is given by
\begin{align}\label{Tau}
\mathcal{T} =& \frac{1}{24}\sum_{acde}\partial_a\partial_c\partial_d\partial_e 
T(\bm{b}_j, \bbeta) \bar{b}_a\bar{b}_c\bar{b}_d\bar{b}_e \nn\\
&+\frac{\ii}{72}\left(\sum_{acd} \partial_a\partial_c\partial_d T(\bm{b}_j, \bbeta) \bar{b}_a\bar{b}_c\bar{b}_d\right)^2\,.
\end{align}
Let us focus first on the GO part of the integral (\ref{FFF}). The resulting amplification factor is given by 
\be
F_{\rm GO}(\omega, \bbeta)= \frac{\bar D_{\rm S} }{\bar D_{\rm L}\bar D_{\rm LS}}\frac{\omega}{2\pi \ii}\sum_j{\rm e}^{\ii\omega T_j}\int \dd^2 \boldsymbol{\bar b}\exp{\left[\frac{\ii\omega}{2\bar L}\boldsymbol{\bar b}^T\tilde H_j\boldsymbol{\bar b}\right]} \,,
\ee
where we defined the dimensionless Hessian $\tilde H_j\equiv \bar L H_j$ (also called an amplification matrix).
Evaluating the 2D Gaussian integral,\footnote{We use $\int_{-\infty}^\infty  {\rm e}^{\pm \ii x^2/2} \dd x= \sqrt{2 \pi} {\rm e}^{\pm \ii \pi/4}$.} we obtain
\be
\int\dd^2 \boldsymbol{\bar b}\exp{\left[\frac{\ii\omega}{2 \bar{L}}\boldsymbol{\bar b}^T\tilde H_j\boldsymbol{\bar b}\right]}=\frac{2\pi \bar{L}}{\omega}\frac{{\rm e}^{\ii\pi/4\sum_a{\rm sgn}(\lambda_a)}}{\sqrt{|\det{\tilde H_j}|}}\,,
\ee
where $\lambda_a$ are the eigenvalues of the Hessian. After summing over all extrema $j$, we obtain 
\be\label{FstGO}
F_{\rm GO}(\omega, \bbeta)= \sum_j \sqrt{\mu_j}  e^{\ii \omega T_j}.
\ee
Here $|\mu_j|$ is the magnification of the $j$-th image defined as $|\mu_j|=|\det{(\bar H_j)}|^{-1}$ and it is combined with the \emph{Morse phase} $n_j\equiv 1/4(2-\sum_a{\rm sgn}(\lambda_a))$ to define
\be\label{Defsqrtmu}
\sqrt{\mu_j} \equiv \sqrt{|\mu_j|}e^{- \ii \pi n_j}\,.
\ee
The Morse phase takes values $n_j=0, 1/2, 1$ when $\bm{b}_j$ corresponds to a minimum, saddle point, or maximum of the time delay function, and equals half the number of negative eigenvalues. Said differently, each negative eigenvalue brings a factor $1/\ii$. Note that, with the definition~\eqref{Defsqrtmu}, $\sqrt{\mu_j}<0$ if $n_j=1$, i.e., for a fully inverted image. If we choose coordinates in the integration plane such that $\partial_{xy} \psi =0$, then 
\be\label{sqrtmuddpsi}
\sqrt{\mu_j} = \frac{1}{\sqrt{1 - \bar{L}\partial_{xx} \psi}} \frac{1}{\sqrt{1 - \bar{L}\partial_{yy} \psi}}\,,
\ee
where $1/\sqrt{-1} = 1/\ii$ if the argument is negative, ensuring the Morse phase is accounted for.

Adding the BE contribution (\ref{Tau}), the final form of the amplification factor is 
\begin{align}\label{Fst}
F(\omega, \bbeta)= \sum_j \sqrt{\mu_j} &\left(1+\frac{\ii}{\omega}\Delta^{\text{st}}_j\right) e^{\ii \omega T_j} +\mathcal{O} \left(\omega^{-2}\right)\,,
\end{align}
where the \emph{st} label denotes the standard contribution coming from the diffraction integral.  The term $\Delta^{\text{st}}_j$ is real and provides a frequency-dependent correction to the GO time delay. At linear order in the lens mass, this reads
\be\label{Delta2}
\Delta^{\text{st}}_j=\frac{1}{8}\left( \frac{\bar D_{\rm L} \bar D_{\rm LS}}{\bar D_{\rm S}}\right)^2\nabla_\perp^4 \psi=\pi G \left( \frac{\bar D_{\rm L} \bar D_{\rm LS}}{\bar D_{\rm S}}\right)^2 \nabla^2_\perp \Sigma\,,
\ee
where we use $\nabla^2_\perp \psi=8\pi G \Sigma$ and
\begin{align}
&\Sigma = \int \rho_{\rm mat}(s){\rm d}s\,,
& \nabla^2_\perp \Sigma= \int \nabla^2_\perp\rho_{\rm mat}(s){\rm d}s\,,
\end{align}
where $s$ is an affine parameter along the GO geodesic.
In polar coordinates for an axisymmetric lens, the phase correction at linear order in the gravitational potential reduces to 
\begin{align}
&\Delta^{\text{st}}_j=\frac{1}{8}\left( \frac{\bar D_{\rm L} \bar D_{\rm LS}}{\bar D_{\rm S}}\right)^2\left[ \psi_j^{(4)}+ \frac{2\psi_j^{(3)}}{|\bm{b}_j|}-\frac{\psi_j''}{|\bm{b}_j|^2}+\frac{\psi_j'}{|\bm{b}_j|^3} \right]\,,
\end{align}
in agreement with the weak-field limit of Eq.\,(13) of \cite{Takahashi:2004mc} after matching the notation. 

Eq.~(\ref{Fst}) shows that, at leading order in the BE expansion, the diffraction integral is dominated by contributions from geometric-optics trajectories. However, the time delay receives a frequency-dependent correction due to diffraction, encoding the interaction of each geometric-optics ray with its local neighbourhood.

\subsection{New ideas}

Instead of relying on the standard assumptions detailed above, our strategy for improving the evaluation of the Kirchhoff integral is to evolve the field from the source to the integration plane using BE equations. Since by construction this approach only allows to describe the field in a neighbourhood of the GO geodesic, we restrict the Kirchhoff integral to a local integration around the extrema as was done in section~\ref{StF}. By doing so, we find that the amplification factor receives a frequency-dependent correction to the time delay arising from amplitude variation across the lens plane and omitted projection effects. In the next section, we summarize the BE formalism developed in \cite{Bruyere:2026gnt}. 

\section{Beyond-eikonal expansion in a nutshell}\label{previous}

The propagation of scalar waves governed by the Klein-Gordon equation in curved spacetime was studied in~\cite{Bruyere:2026gnt}, focusing on corrections to the wave amplitude and phase. Using the Newman-Penrose formalism~\cite{Newman:1961qr}, the first BE corrections were derived in an inverse-frequency expansion, building on ideas developed in Refs.~\cite{MTW,Dolan:2017zgu,Dolan:2018ydp,Harte:2018wni,Harte:2019tid, Cusin:2019rmt, Dalang:2021qhu,  Menadeo:2025hgf}. It was shown that in vacuum, where lensing arises from the Weyl tensor, wave effects vanish at linear order in the gravitational potential $\phi$ and first appear at order $\phi^2$, a result differing from that obtained via partial wave scattering resummation in~\cite{CarrilloGonzalez:2025gqm}. In contrast, for propagation through matter (with non-vanishing density), the leading corrections appear at order $\phi$.

A geometric-optics trajectory receives corrections from its neighbourhood, defined as a region of size $a\simeq \sqrt{\lambda L}$, where 
\be\label{DefL}
L=D_{\rm L}D_{\rm LS}/D_{\rm S}\quad \Leftrightarrow \quad \frac{1}{L} = \frac{1}{D_{\rm LS}} + \frac{1}{D_{\rm L}}\,. 
\ee
This relation is the finite-distance analogue of Eq.~\eqref{def_L}. 
The length $a$ corresponds to the size of a Fresnel zone (see e.g. section 8.2 of \cite{BornWolf:1999:Book} or~\cite{Leung:2023lmq}), that is to a beam region in which the phase difference due to the variation of the propagating length is of order unity.
The perturbative approach of \cite{Bruyere:2026gnt} is valid in the weak-field limit, for $b>a$, where $b$ is the characteristic scale of the lens potential. Hence, the small parameter governing the size of BE corrections for a point-like lens is $\omega_{\rm ref}/\omega$ where $\omega_{\rm ref} \equiv L/b^2$.

\subsection{Newman-Penrose scalars}

The Newman-Penrose (NP) formalism uses (complex valued) scalars only, by projecting all tensors onto a null tetrad.
It is convenient to choose a parallel transported null tetrad $\{k^a,n^a,m^a,\bar m^a\}$, where the vector $\bm{k}$ is identified with the geometric-optics wave vector. First, we define twelve complex spin scalars describing the affine connection, which are
\begin{align}\label{NP}
& \rho \equiv - m^a\bar{m}^b\nabla_bk_a \,, \hspace{0,3cm} \lambda\equiv\bar{m}^a\bar{m}^b\nabla_bn_a \,, \\
& \sigma\equiv-m^am^b\nabla_bk_a \,, \hspace{0,3cm} \mu\equiv\bar{m}^am^b\nabla_bn_a \,, \\
& \tau\equiv-m^an^b\nabla_bk_a,\hspace{0,5cm} \nu\equiv\bar{m}^an^b\nabla_bn_a \,, \\
& \beta\equiv-\frac{1}{2}(n^am^b\nabla_bk_a-\bar{m}^am^b\nabla_bm_a)\,, \\
& \gamma\equiv-\frac{1}{2}(n^an^b\nabla_bk_a-\bar{m}^an^b\nabla_bm_a)\,.
\end{align}
Curvature in vacuum is described by five complex scalars encoding the Weyl projections, defined by
\begin{align}\label{DefPsin}
 & \Psi_0\equiv C_{abcd}k^am^bk^cm^d \,,\,\, \Psi_1\equiv C_{abcd}k^an^bk^cm^d \,, \\
 & \Psi_2\equiv C_{abcd}k^am^b\bar{m}^cn^d  \,,\,\, \Psi_3\equiv C_{abcd}k^an^b\bar{m}^cn^d \,, \\
 & \Psi_4\equiv C_{abcd}n^a\bar{m}^bn^c\bar{m}^d \,,
\end{align}
where $C_{abcd}$ is the Weyl tensor. When matter is present, we must also consider the components of the Ricci tensor. These ten degrees of freedom are organized as four real valued and three complex scalars, which are
\begin{align}
& \Phi_{00}\equiv\frac{1}{2}R_{ab}k^ak^b \,,\,\, \Phi_{11}\equiv\frac{1}{4}R_{ab}(k^an^b+m^a\bar{m}^b) \,, \\
& \Phi_{22}\equiv\frac{1}{2}R_{ab}n^an^b \,,\,\, \Lambda\equiv\frac{R}{24} \,,\,\, \Phi_{01}\equiv\frac{1}{2}R_{ab}k^am^b =\bar{\Phi}_{10} \, \\
& \Phi_{02}\equiv\frac{1}{2}R_{ab}m^am^b =\bar{\Phi}_{20} \,, \hspace{0,2cm} \Phi_{12}\equiv\frac{1}{2}R_{ab}m^an^b =\bar{\Phi}_{21}.
\end{align}
Finally, we introduce the projection of covariant derivatives on the null tetrad. 
\begin{align}
& D\equiv k^a\nabla_a \,\,,\,\,\, \Delta_d\equiv n^a\nabla_a \,\,,\,\,\, \delta\equiv m^a\nabla_a \,\,,\,\,\, \bar\delta\equiv \bar m^a\nabla_a \,\,.
\end{align}
$D$ is the derivative along the corresponding geodesic, i.e., $D=\dd/\dd s$, where $s$ is the affine parameter. The NP formalism allows us to write the lensing equations as a set of coupled equations involving scalar quantities only, by projecting tensorial quantities onto the tetrad and introducing the corresponding scalars (spin, curvature, or covariant-derivative scalars).

\subsection{Perturbative approach}

In full generality, BE corrections can be expressed as complex corrections to the amplitude. At first order in the BE expansion, we write~\cite{Harte:2018wni,Dolan:2018ydp,Menadeo:2025hgf}
\be
\hat{\Psi}=A_0\left(1+\omega^{-1}\frac{A_1}{A_0}+\cdots\right){\rm e}^{\ii\omega S} \,,
\ee
where $\nabla_{a} S=k_{a}$. 
In \cite{Bruyere:2026gnt}, the full system of coupled differential equations governing the NP scalars and the two amplitudes $A_0$ and $A_1$ was obtained. First, the transport equation for the wave amplitude reads \cite{Bruyere:2026gnt} 
\begin{align}
&2 D A_0+A_0 \nabla_{a}k^{a}=0\,,\label{evolA0}\\
&D \left(  \frac{A_1}{A_0}\right)=\frac{\ii}{2}\frac{\Box A_{0}}{A_0}\,.\label{evolA1}
\end{align}
We may always choose $A_0$ real; therefore, from the above equation $A_1$ is purely imaginary.

\subsection{Eikonal evolution}\label{eikonal_evolution}

We then exhibit the subsystem of equations that must be solved to recover the GO form for the field. The GO amplitude $A_0$, the beam expansion rate $\rho$, and the beam expansion shear $\sigma$ satisfy
\begin{align}
&D A_0=A_0 \rho \label{eqA0}\,,\\
&D\rho= \rho^2 +\sigma \bar{\sigma}+\Phi_{00} \label{eqrho}\,,\\
&D\sigma=2 \rho\sigma +\Psi_0 \label{eqsig}\,,
\end{align}
where the first equation is follows from Eq.~\eqref{evolA0} using that $\rho=-1/2 \nabla_a k^a$ (in the absence of twist). The deformation rate matrix~\cite{Perlick:2010zh,Fleury:2015hgz} can be constructed from $\rho$ and $\sigma$; it is related to the Jacobi matrix, which describes the shape of an infinitesimal beam of geodesics (see e.g. Appendix E of \cite{Bruyere:2026gnt}). One can then show that $\rho=-1/(2\mathcal{A})D\mathcal{A}$ where $\mathcal{A}$ is the cross-sectional area of the ray bundle, hence $A_0 \propto 1/\sqrt{\mathcal{A}}$.
Since the magnification $\sqrt{\mu}$ is defined as 
\be
\sqrt{\mu}\equiv \frac{A_0}{A_0^{\rm NL}} \,,
\ee
where $A_{\rm NL}$ is the unlensed amplitude, then $\mathcal{A} = \mathcal{A}_{\rm NL} /\sqrt{\mu}^2$, where $\mathcal{A}_{\rm NL}$ is the unlensed cross-sectional area. The amplification factor at the GO order is then related to the magnification via
\be
F_{\rm GO} \equiv \frac{ A_0 {\rm e}^{\ii \varphi} }{ A_0^{\rm NL} {\rm e}^{\ii \varphi_{\rm NL}}} = \sqrt{\mu} \,{\rm e}^{\ii \omega T}\,,
\ee
in agreement with each term of~\eqref{FstGO}.

To determine the magnification, we introduce an affine parameter $s$ that vanishes at the source. In particular, it coincides with the distance along the geodesic, namely $D_{\rm LS} = s_L$, $D_{\rm L} = s-s_L$ and $D_{\rm S} = s$. Therefore, $L(s)$ defined in~\eqref{DefL} can be equivalently written as
\be
L(s)=\frac{s_{\rm L}(s-s_{\rm L})}{s} \,.
\ee
Under the thin lens approximation for a beam passing through a lens at $s=s_L$, the solutions after the lens for the system~(\ref{eqA0}-\ref{eqsig}) are (see e.g. section 3.3 of \cite{Bruyere:2026gnt}) 
\begin{align}\label{rhoThin}
\rho =& -\frac{1}{2 s}[1-s_L (\mathcal{I}_{\Phi_{00}} - |\mathcal{I}_{\Psi_0}|)] {\cal M}^2_{\mathcal{I}_{\Phi_{00}} - |\mathcal{I}_{\Psi_0}|}(s)\nonumber\\
&-\frac{1}{2s}[1-s_L (\mathcal{I}_{\Phi_{00}} + |\mathcal{I}_{\Psi_0}|)] {\cal M}^2_{\mathcal{I}_{\Phi_{00}} + |\mathcal{I}_{\Psi_0}|}(s)\,,
\end{align}
\begin{align}\label{sigThin}
|\sigma|=&\frac{1}{2s}[1-s_L (\mathcal{I}_{\Phi_{00}} - |\mathcal{I}_{\Psi_0}|)] {\cal M}^2_{\mathcal{I}_{\Phi_{00}} - |\mathcal{I}_{\Psi_0}|}(s)\nonumber\\
&-\frac{1}{2s}[1-s_L (\mathcal{I}_{\Phi_{00}} + |\mathcal{I}_{\Psi_0}|)] {\cal M}^2_{\mathcal{I}_{\Phi_{00}} + |\mathcal{I}_{\Psi_0}|}(s)\,,
\end{align}
where we defined
\be
{\cal M}_{Q}(s) \equiv \frac{1}{\sqrt{1 - L(s) Q}}\,,
\ee
and $\mathcal{I}_{\Psi_0}\equiv\int\Psi_0\dd s$, $\mathcal{I}_{\Phi_{00}} \equiv \int \Phi_{00} \dd s$. The solution for the amplitude is 
\begin{align}\label{A0Thin}
A_0(s) = \frac{A_0(s_L) s_L}{s} \, {\cal M}_{\mathcal{I}_{\Phi_{00}} + |\mathcal{I}_{\Psi_0}|}(s)\, {\cal M}_{\mathcal{I}_{\Phi_{00}} - |\mathcal{I}_{\Psi_0}|}(s)\,,
\end{align}
and therefore
\be\label{mubeautiful}
\sqrt{\mu}(s) = {\cal M}_{\mathcal{I}_{\Phi_{00}} + |\mathcal{I}_{\Psi_0}|}(s)\, {\cal M}_{\mathcal{I}_{\Phi_{00}} - |\mathcal{I}_{\Psi_0}|}(s)\,.
\ee

When a caustic is crossed, one or both factors $1 - L(s) (\mathcal{I}_{\Phi_{00}} \pm |\mathcal{I}_{\Psi_0}|)$ change sign, and the corresponding ${\cal M}_{\mathcal{I}_{\Phi_{00}} \pm |\mathcal{I}_{\Psi_0}|}$ become $\propto 1/\sqrt{-1} = 1/\ii $. However, since the differential system goes through a discontinuity, nothing guarantees that this is the correct solution after caustic crossing. For instance~\eqref{A0Thin} could be replaced by 
\be
A_0(s) = \frac{A_0(s_L) s_L/s}{\sqrt{\prod_{r = \pm 1}[1 - L(s)(\mathcal{I}_{\Phi_{00}} + r |\mathcal{I}_{\Psi_0}|)]}}\,,
\ee
which is equal to~\eqref{A0Thin} before caustic crossing, but differs by a phase when both factors have changed sign. One could even consider the solution $\propto \sqrt{1/\prod_{r = \pm 1}[1 - L(s)(\mathcal{I}_{\Phi_{00}} + r |\mathcal{I}_{\Psi_0}|)]}$ which would have a sign difference after one of the factors has changed sign.

This phase ambiguity, mathematically associated with the branch cut of the square root in the complex plane, is related to the Morse phase introduced in sections~\ref{StF} and~\ref{SecGO}. Since the GO determines the phase via a local Hamilton–Jacobi equation and yields a single-valued eikonal function $S$, it cannot correctly describe caustics where geodesics intersect. Only a global wave-optical treatment, e.g. with the Kirchhoff integral, can unambiguously fix the phase after caustic crossing.

Finally if we choose the coordinates system such that $\partial_{xy} \psi = 0$, then with results of section 2.11 of \cite{Bruyere:2026gnt}, $\mathcal{I}_{\Phi_{00}} = 1/2(\partial_{xx}\psi + \partial_{yy}\psi)$ and $\mathcal{I}_{\Psi_0} = 1/2(\partial_{xx}\psi - \partial_{yy}\psi)$, and we check that the magnification~\eqref{mubeautiful} is similar to~\eqref{sqrtmuddpsi} with $\bar D_{\rm L}\bar D_{\rm LS}/\bar D_{\rm S} \to s_L (s-s_L)/s = D_{\rm LS}D_{\rm L}/D_{\rm S}$,  that is $\bar{L} \to L$.

\subsection{Beyond-eikonal evolution}

The NP scalars \eqref{NP} obey evolution equations, e.g. \eqref{eqrho} and \eqref{eqsig} governing the evolution of $\rho$ and $\sigma$.
To solve Eq.~\eqref{evolA1} giving the evolution of the BE amplitude correction, we need $\Box A_0$ to compute the transport of $A_1$ along the ray. This requires knowing, along the geodesic, terms of the form $\nabla_a \nabla_b A_0$, projected onto the tetrad vectors. In other words, one rewrites $\Box A_0$ as a combination of $A_0$ and its projected derivatives $\delta$ and $\bar{\delta}$, with coefficients given by spin scalars, see section 2.6 of \cite{Bruyere:2026gnt}. Thus, we also need the evolution of the projected derivatives of spin scalars which are also reported in \cite{Bruyere:2026gnt}. The right-hand side of Eq.~\eqref{evolA1} reads
\begin{align}\label{BoxA0}
\Box A_0=&\frac{\delta\bar\delta A_0}{A_0}-\delta\bar\tau+(\bar\mu-\mu)\rho+2\Lambda+\Psi_2 \\
& +\sigma\lambda-2(\bar\beta-\bar\tau)\tau+2(\beta-\tau)\frac{\bar\delta A_0}{A_0}-\bar\tau\frac{\delta A_0}{A_0} \nn\,.
\end{align}
It is then clear that to solve for the amplitude $A_1$, we must solve a system of coupled differential equations for all spin scalars and their projected derivatives in~\eqref{BoxA0}.  In \cite{Bruyere:2026gnt} the resulting system of coupled differential equations is solved analytically and numerically for different lens models. The analytical results were obtained by linearising in $\phi$, and by considering the thin lens approximation. In contrast, no approximations were used in the numerical solutions. 

\subsection{Phase correction}\label{phasecorrection}

Since $A_1$ is purely imaginary, it contributes a phase correction to the field at the observer. The phase of the wave receives BE corrections of the form 
\begin{align}\label{DefDelta1}
&S\rightarrow S+\omega^{-2}\Delta\,,
&\Delta=\frac{\Im(A_1)}{A_0} \,.
\end{align}
For a matter lens, it was found in \cite{Bruyere:2026gnt} at first order in $\phi$ and with the thin lens approximation that
\be\label{MainRes}
\Delta= \Delta^{\text{st}}+\Delta^{\text{new}}\,,
\ee
where the standard diffraction integral contribution and the new contribution are
\begin{align}
&\Delta^{\text{st}}=  
\pi G\nabla^2_{\perp}\Sigma\frac{(u_{\rm L}-u)^2}{u_{\rm L}^4}\,,\label{varphist}\\
&\Delta^{\text{new}}=2\pi G \Sigma  \frac{u^2}{u_{\rm L}^2}\,, \label{varphinew}
\end{align}
with  
\be
u\equiv \frac{1}{s}\,.
\ee
Here $\nabla_{\perp}^2\equiv  2\delta \bar{\delta}$ corresponds to the Laplacian on the polarization plane.\footnote{It corresponds to the two-dimensional Laplacian in the flat plane spanned by linear combinations of $\bm m$ and $\bm{\bar{m}}$. When acting on the integrated surface density $\Sigma$, it is the Laplacian in the lens plane.} The L subscript indicates that variables are evaluated on the lens plane. Eqs.~\eqref{varphist} and \eqref{varphinew} can be rewritten in terms of physical distances as
\begin{align}\label{2Delta}
&\Delta^{\text{st}}= \pi G \nabla^2_{\perp} \Sigma\left( \frac{D_{\rm L} D_{\rm LS}}{D_{\rm S}}\right)^2\,,\\
&\Delta^{\text{new}}= 2\pi G \Sigma  \frac{D_{\rm LS}^2}{D_{\rm S}^2}\,.
\end{align}
For each geometric-optics image, there are two differences from the standard diffraction integral result, see Eqs.~\eqref{Fst}. First, there is an additional correction; second, all distances are evaluated along the GO geodesic and reduce to standard (barred) distances only if the transverse lensing components are negligible, which is the underlying assumption of the diffraction integral.\footnote{We stress that the difference between geodesic distances and bar distances is of of order of the correction $\Delta^{\rm new}$, hence this cannot be ignored for internal consistency. In other terms, since we focus on corrections which are of order $(b/D)^2$ we have to put the correct distances (along geodesic distances) in the dominant/standard term.} The standard contribution corresponds to the result found with the diffraction integral \eqref{Delta2} but with refined distances (this amounts to a subdominant correction), around a given GO solution $j$ (here the label $j$ is omitted because we are following a given trajectory).

The new term is suppressed by $(b/D_{\rm L})^2$ relative to the standard term. As observed  in \cite{Bruyere:2026gnt}, just after the thin lens, the \emph{new} term \eqref{varphinew} dominates and is proportional to the lens surface density $\Sigma$. This contribution fades as $(D_{\rm L}/D_{\rm S})^2$. At large distances from the lens, the \emph{standard} contribution (\ref{varphist}) dominates and is proportional to $\nabla_{\perp}^2\Sigma$. As discussed, this formulation captures
diffraction effects in a neighborhood of the geometric
optics trajectory.

\section{Improved diffraction integral}\label{improved}

We now aim to evaluate the Kirchhoff integral \eqref{surfText}, relaxing the standard approximations that led to the diffraction integral \eqref{finalSt}. We follow the procedure outlined below (see right panel of Fig.~\ref{Lens0} for a schematic illustration of the procedure followed): 
\begin{enumerate}
    \item we propagate the field from the source to the integration plane (which is an arbitrary plane after the lens plane) using BE evolution, following the procedure of \cite{Bruyere:2026gnt} summarized in the previous section,
     \item we evaluate locally the Kirchhoff integral in order to account for diffraction effects, choosing an integrating plane orthogonal to the geometric-optics trajectory, at or after the lens plane,
     \item finally, we sum the contributions associated with all geometric-optics trajectories.
\end{enumerate}
Crucially, we will show that the final result is independent of the choice of the integration plane beyond the lens (see the end of Sec.~\ref{linear}). This allows us to switch at any stage between the BE expansion and the Kirchhoff integral to compute the field measured by the observer, thereby establishing the equivalence of the methods.

\subsection{Choice of coordinates}

We choose an integration plane located after the lens plane and denote by K the point where it intersects the GO geodesic. We then introduce a coordinate system with the $z$ axis along the wave-vector (see Fig.~\ref{setup}). We define $D_{\rm K}$ as the distance between K and O, and $D_{\rm KS}$ as the distance between S and K, both measured along the geodesic path. We use the coordinates ($x$, $y$) in the vicinity of the geodesic and introduce the complex notation
\be
\zeta = \frac{1}{\sqrt{2}}(x+\ii y) \quad\Rightarrow\quad 2\zeta\bar{\zeta} = x^2 + y^2 \,.
\ee
In this coordinates system, $\ell$ is given by
\be\label{lexpr}
\ell= \sqrt{(D_{\rm K}-z)^2 + 2 \zeta \bar{\zeta} } \,.
\ee

\subsection{General expression in weak-field approximation}\label{wf}

We evaluate the Kirchhoff integral \eqref{surfText} by expanding the integrand around each GO trajectory. We denote quantities associated with the $j$-th image with the superscript $(j)$. However, to avoid heavy notation, we omit this superscript throughout this section unless there is a risk of ambiguity. It will be restored in the final expression for the field at the observer. In particular, all distances from the lens appearing in this section are implicitly understood to correspond to the $j$-th image.

We Taylor expand the amplitude $A$ and the phase $\varphi$ of the field on the lens plane as 
\begin{align}
&A\simeq A\big|_{\rm K} + \left.x^a (\nabla_a A)\right|_{\rm K} + \frac{1}{2}\left. x^a x^b(\nabla_a\nabla_b A)\right|_{\rm K} +\cdots \,, \label{expAmp}
\end{align}
\begin{align}
\varphi\simeq \varphi\big\vert_{\rm K} &+\left.k_a x^a\right|_{\rm K}+\frac{1}{2}\left.x^b x^a(\nabla_a k_b)\right|_{\rm K} \nn\\
& + \frac{1}{6} \left.x^a x^b x^c(\nabla_a \nabla_b k_c)\right|_{\rm K} \nn\\
& + \frac{1}{24} \left.x^a x^b x^c x^d(\nabla_a \nabla_b \nabla_c k_d)\right|_{\rm K} +\cdots \,,\label{expPhase}
\end{align}
with 
\begin{align}
x^a =\bar\zeta m^a+\zeta\bar m^a + \frac{z}{2}(k^a-2n^a) \,.
\end{align}
In particular, we have $k_a x^a = z$. Note that $x^a$ here denotes coordinates on a plane orthogonal to the geometric-optics trajectory (see Fig.\,\ref{setup}), whereas in the standard evaluation of the diffraction integral BE one expands around the geometric-optics trajectory on a plane orthogonal to the optical axis; see Eq.\,\eqref{tdex}. However, these two approaches are equivalent. This follows from the fact that the choice of the integration surface in the Kirchhoff formalism is arbitrary; see Appendix~\ref{Kirchhoff} for details.

We shall now consider a weak-field approximation and we expand $\Psi$ in orders of the gravitational potential $\phi$. Only quadratic terms in $\zeta,\bar\zeta$ are kept in the phase so that the expansion is the product of a polynomial in $\zeta,\bar{\zeta}$ with a two-dimensional Gaussian function. As a consequence we can keep only even powers in $\zeta,\bar\zeta$ in this polynomial and discard odd terms in $\zeta,\bar\zeta$. Moreover since we integrate on the surface at $z=0$ but there is a single derivative with respect to $z$, the terms with powers of $z$ larger than one will always vanish. We also only keep for the moment second order contributions in $\phi$ and we shall later restrict to a first order only expansion.

We explicitly compute \eqref{expAmp} and \eqref{expPhase} by expressing covariant derivatives of $k_a$ in terms of the NP scalars and the tetrad vectors, using the strategy from \cite{Bruyere:2026gnt}. We contract covariant derivatives with tetrad vectors and then commute them to obtain a result in terms of only spin scalars and projected derivatives (and we reiterate the procedure when needed). For the phase expansion, we need 
\begin{align}
& x^b x^a(\nabla_a k_b) \supset 2\bar\tau z\zeta + 2\tau z\bar\zeta - 2 \rho \zeta \bar{\zeta} - \sigma \bar{\zeta}^2 - \bar{\sigma} \zeta^2 \,, \label{exp1}\\
& x^a x^b x^c(\nabla_a \nabla_b k_c) \supset \mathcal{K}_1 z\zeta\bar\zeta + \mathcal{K}_2 \zeta^3 + \mathcal{K}_3 \bar\zeta^3 +\mathcal{K}_4\zeta\bar\zeta^2 \nn \\
&\hspace{3cm} +\mathcal{K}_5\zeta^2\bar\zeta \,, \\
& x^a x^b x^c x^d(\nabla_a \nabla_b \nabla_c k_d) \supset \mathcal{K}_6 \zeta^2 \bar \zeta^2 \,,
\end{align}
where the symbol $\supset$ indicates that only terms up to second order in the potential that give non-vanishing contributions are kept. Similarly, the amplitude \eqref{expAmp} is expanded as 
\begin{align}
A &\simeq A_0+ \frac{A_1}{\omega} + z A_0 \rho + \mathcal{K}_7\,\zeta\bar\zeta\,.
\end{align}
The explicit forms of the coefficients $\mathcal{K}_1\,,\dots\,,\mathcal{K}_7$ in terms of spin scalars evaluated at K are gathered in Appendix~\ref{coeff}. 

Eventually we obtain 
\begin{align}
&\left.\Psi\right|_{z=0} \supset \Big(A_0+ \frac{A_1}{\omega} + \mathcal{K}_7 \zeta \bar\zeta + \ii\frac{A_0}{24} \mathcal{K}_6 \omega\zeta^2\bar\zeta^2 \nn\\
& \hspace{1,8cm} -\frac{A_0}{36} \left(\mathcal{K}_2\mathcal{K}_3+\mathcal{K}_4\mathcal{K}_5\right) \omega^2\zeta^3\bar\zeta^3 \Big) \e^{-\ii\omega\rho\zeta\bar\zeta} \nn \\
&\hspace{1,8cm} \times \e^{-\ii\omega(\sigma \bar{\zeta}^2 + \bar{\sigma} \zeta^2)/2} \,\e^{\ii\omega\varphi}\,,
\end{align}
and
\begin{align}
&\left.\partial_z\Psi\right|_{z=0} \supset \ii\omega\left( A_0 + \frac{A_1}{\omega} -\frac{\ii\rho A_0}{\omega} + \mathcal{K}_7 \zeta \bar\zeta +\frac{A_0}{6} \mathcal{K}_1 \zeta\bar\zeta \right.\nn\\
& \hspace{1,3cm} \left. + \ii\frac{A_0}{24} \mathcal{K}_6 \omega\zeta^2\bar\zeta^2 -\frac{A_0}{36} \left(\mathcal{K}_2\mathcal{K}_3+\mathcal{K}_4\mathcal{K}_5\right) \omega^2\zeta^3\bar\zeta^3 \right) \nn\\
& \hspace{2cm} \times \e^{-\ii\omega\rho\zeta\bar\zeta}\e^{-\ii\omega(\sigma \bar{\zeta}^2 + \bar{\sigma} \zeta^2)/2}\,\e^{\ii\omega\varphi} \,.
\end{align}
We also need to expand the Green function. Using~\eqref{lexpr}, and expanding around K, we get
\begin{align}\label{exp}
\left.\frac{\e^{\ii\omega \ell}}{\ell}\right|_{z=0} \simeq \left( 1-\frac{\zeta\bar\zeta}{D_{\rm K}^{2}}-\frac{\ii\omega}{2}\frac{\zeta^2\bar\zeta^2}{D_{\rm K}^{3}} \right) \frac{\e^{\ii\omega D_{\rm K}}\e^{\ii\omega\zeta\bar\zeta/D_{\rm K}}
}{D_{\rm K}}\,,
\end{align}
and its derivative is
\begin{align}\label{dexp}
&\left. \partial_z \left( \frac{\e^{\ii\omega \ell}}{\ell}\right)\right|_{z=0} \simeq\nn\\
&-\ii\omega\left( 1-\frac{2\zeta\bar\zeta}{D_{\rm K}^{2}} +  \frac{\ii}{\omega D_{\rm K}} -\frac{\ii\omega}{2}\frac{\zeta^2\bar\zeta^2}{D_{\rm K}^{3}} \right) \frac{\e^{\ii\omega D_{\rm K}}\e^{\ii\omega\zeta\bar\zeta/D_{\rm K}}}{D_{\rm K}} \,.
\end{align}

We can now write the field at the observer \eqref{surfText}. We put the index $(j)$ on the left hand side to refer to the $j$-th image. All quantities on the right also refer to the $j$-th image but we omit the index to alleviate the notation. We obtain the central result of this article
\begin{widetext}
\begin{align} \label{wave}
& \Psi^{(j)} = \frac{\omega}{2\ii \pi}\frac{\e^{\ii\omega\varphi_{\rm GO}}}{D_{\rm K}}\bigintsss_{\rm K}\dd\zeta \dd\bar{\zeta}\,\, A_0\Bigg[ 1 + \left( \frac{A_1}{A_0} -\frac{\ii}{2}\rho + \frac{\ii}{2}\frac{1}{D_{\rm K}} \right)\frac{1}{\omega} + \left( \frac{\mathcal{K}_7}{A_0} + \frac{\mathcal{K}_1}{12} - \frac{3}{2}\frac{1}{D_{\rm K}^{2}} \right)\zeta\bar\zeta \nn \\
&\hspace{3cm} + \frac{\ii}{2} \left( \frac{\mathcal{K}_6}{12} -\frac{1}{D_{\rm K}^{3}} \right)\omega\zeta^2\bar\zeta^2 - \frac{1}{36} \left(\mathcal{K}_2\mathcal{K}_3+\mathcal{K}_4\mathcal{K}_5\right) \omega^2\zeta^3\bar\zeta^3 \Bigg] \e^{\ii\omega\zeta\bar\zeta\left(1/D_{\rm K}-\rho\right)} \e^{-\ii\omega(\sigma \bar{\zeta}^2 + \bar{\sigma} \zeta^2)/2} \,,
\end{align}
\end{widetext}
where the GO phase $\varphi_{\rm GO}$ is given by 
\be\label{phiGO}
\varphi_{\rm GO}= D_{\rm S}^{\rm str}+T \,,
\ee
with $T$ the time delay defined in \eqref{timed}.  
We recall that all quantities are evaluated on the central GO trajectory. To evaluate this integral, one also needs to know the optical scalars on the lens plane, as well as $A_0$ and $A_1$ that characterize the field on the lens plane. 

We will show that once the integral is performed, the final result no longer depends on the point $K$, i.e., on the choice of the surface where we perform the Kirchhoff integral, thereby proving that the BE formalism of section~\ref{previous} is equivalent to our improved Kirchhoff integration method. In the next subsection we will first show this property for the leading order in $1/\omega$ which corresponds to the GO optics. Then section~\ref{SecGauss} is dedicated to the computation of the first BE corrections.
 
\subsection{Geometric-optics limit}\label{SecGO}

We now check that the GO limit of \eqref{wave}, which corresponds to $\omega\rightarrow\infty$, reduces to the standard result after summing over all images. Therefore, in \eqref{wave} we keep only the term that is not of order $1/\omega$. For a specific image trajectory, we obtain
\begin{align}\label{psiGO}
\Psi^{(j)}_{\rm GO} =& \frac{\omega}{2\ii \pi}\frac{\e^{\ii\omega \varphi_{\rm GO}}}{D_{\rm K}}\bigintsss_{\rm K}\dd\zeta \dd\bar{\zeta}\,\, A_0 \exp\left[\ii\omega\zeta\bar\zeta\left(\frac{1}{D_{\rm K}}-\rho\right)\right] \nn\\
&\hspace{2,cm} \times\exp\left[-\frac{\ii \omega}{2}\left(\sigma \bar{\zeta}^2 + \bar{\sigma} \zeta^2\right) \right] .
\end{align}
We can compute this integral using the same tools as in Sec.~\ref{StF}. The dimensionless Hessian matrix for this Gaussian integral is
\be\label{defHbarab2}
\bar H= L_{\rm K}
\begin{pmatrix}
1/D_{\rm K}-\rho-\Re(\sigma)/2 & -\Im(\sigma)/2  \\
-\Im(\sigma)/2 & 1/D_{\rm K}-\rho+\Re(\sigma)/2
\end{pmatrix}\,,
\ee
where we introduced the length scale
\be\label{L}
L_{\rm K}\equiv \frac{(u_{\rm K}-u)}{u_{\rm K}^2} = \frac{D_{\rm K}D_{\rm KS}}{D_{\rm S}} \,,
\ee
and $D_{\rm K} = s-s_{\rm K}$. When K is chosen on the lens plane, $L_{\rm K}$ reduces to the distance $L$ defined in~\eqref{DefL}. We then use the thin lens results~\eqref{rhoThin} and~\eqref{sigThin}, and, after performing the Gaussian integral, we get
\begin{align}\label{psiGOjraw}
\Psi^{(j)}_{\rm GO} =& \frac{s_{\rm K}}{s}A_0(s_{\rm K}) {\rm e}^{\ii\omega \varphi_{\rm GO}}\nonumber\\
&\times\frac{{\cal M}_{\Sigma - |\mathcal{I}_{\Psi_0}|}(s){\cal M}_{\Sigma + |\mathcal{I}_{\Psi_0}|}(s)}{{\cal M}_{\Sigma - |\mathcal{I}_{\Psi_0}|}(s_{\rm K}){\cal M}_{\Sigma + |\mathcal{I}_{\Psi_0}|}(s_{\rm K})}\,.
\end{align}
Then, substituting the amplitude from~\eqref{A0Thin} evaluated at ${\rm K}$, we obtain
\begin{align}\label{psiGOj}
\Psi^{(j)}_{\rm GO}(s) = \frac{A_0(s_{\rm L}) s_{\rm L}}{s} \sqrt{\mu_j}(s){\rm e}^{\ii\omega \varphi_{\rm GO}}
\end{align}
where $\sqrt{\mu}(s)$ is defined in~\eqref{mubeautiful}. For each image, we recover, as expected, the result of the full BE method, Eq.~\eqref{A0Thin}, but this time without any ambiguity about the phase. More importantly, we have shown that the result does not depend on the point $K$ at which we performed the local Kirchhoff integration. Indeed, once expressed as an amplification factor, it is simply~\eqref{FstGO}. The diffraction integral and our improved Kirchhoff integration lead to the same result when restricted to the GO order. Differences only arise when we consider the first corrections beyond the GO result, that is when we consider corrections of order $1/\omega$. Section~\ref{SecGauss} is devoted to the computation of these corrections. 

Finally let us linearize \eqref{psiGOj} in $\phi$ in order to have a consistent solution when we derive the first order corrections. To that purpose we define $\Psi_{\text{GO}}^{(j),\mu=1}$ as Eq.~\eqref{psiGOj} when $\mu=1$, that is via $\Psi_{\text{GO}}^{(j)} = \Psi_{\text{GO}}^{(j),\mu=1} \sqrt{\mu_j}$.
At first order in $\phi$ we get
\be\label{Psimag}
\Psi^{(j)}_{\rm GO}=\Psi_{\text{GO}}^{(j),\mu=1}\left(1+\Delta^{\rm mag}_j\right) \,.
\ee
where 
\be\label{Dmag}
\Delta^{\rm mag}_j = 4 \pi G \Sigma \frac{D_{\rm L} D_{\rm LS}}{D_{\rm S}} \,.
\ee
This is a GO contribution and must therefore be retained to obtain a consistent solution at linear order in $\phi$ including BE corrections [see e.g.~\eqref{IJ}].

\section{First order corrections in the improved diffraction integral}\label{SecGauss}

We now derive the $1/\omega$ BE corrections via the Kirchhoff integral and show their exact correspondence with the pure BE formalism. We restrict to first order in the gravitational potential $\phi$.\footnote{See section~\ref{NL} for the dominant non-linear contributions for an axisymmetric lens.  We stress that the linear contribution vanishes in vacuum, hence in Appendix~\ref{PointLikeLens} we compute the dominant quadratic contribution in  that case.}

To clarify the origin of the different contributions, we proceed through a sequence of cases of increasing complexity. Rather than applying the full formalism directly, we first isolate the different contributions contained in the Kirchhoff integral. 
We first consider wave propagation in the absence of a lens, on a flat background. We begin with a spherical wave and then a wave with a non-constant amplitude. These simple cases allow us to isolate the different structures contained in the Kirchhoff integral before proceeding to the full derivation where the gravitational potential of the lens is taken into account. 

\subsection{Gaussian integrals}\label{GaussInt}

Let us introduce the tools required to compute the integral~\eqref{wave}.
At linear order in $\phi$, it can be rewritten in a compact form by collecting terms with the same power of $\zeta\bar\zeta$. In this process, we also choose to expand the exponential around its flat spacetime value. The resulting expression takes the general form
\begin{align}\label{gaussian}
\Psi^{(j)}& \equiv A e^{\ii \omega\varphi}=
 e^{\ii\omega\varphi_{\rm GO}}\frac{A_0\vert_{\rm K} L_{\rm K}}{D_{\rm K}} \\
 &\times\int_{\rm K} \dd^2 \boldsymbol{x} \,\bar{g}(\boldsymbol{x})\left[\mathcal{C}_0 + \mathcal{C}_1 \omega \zeta \bar \zeta + \mathcal{C}_2 (\omega\zeta \bar \zeta)^2 + \mathcal{C}_3 (\omega\zeta \bar \zeta)^3\right] , \nn
\end{align}
where $\bar{g}(\boldsymbol{x})$ is the normalized Gaussian function
\be
\bar{g}(\boldsymbol{x}) = \frac{\omega}{2\pi \ii L_{\rm K}}\exp\left(\ii \omega \frac{2\zeta\bar\zeta}{2L_{\rm K}}\right)\,.
\ee
Under the form~\eqref{gaussian}, it becomes clear that we need to compute the following Gaussian integrals (recall that $2\zeta\bar\zeta=x^2+y^2$) 
\be\label{Gint}
\int \dd^2\boldsymbol{x} \,\bar{g}(\boldsymbol{x}) (\omega\zeta \bar \zeta)^n = n! (\ii L_{\rm K})^n\,.
\ee
The next section, is dedicated to the practical evaluation of the integral~\eqref{gaussian}.

\subsection{Spherical wave}\label{sph}

We first consider the case where no lens is present and evaluate the Kirchhoff integral \eqref{wave} by keeping only the background contributions (i.e., zeroth order in the lens potential). 
In flat spacetime, a spherical wave is described by $\left.A_0\right|_{\rm K}=\mathcal{A}/s_{\rm K}$, $\mathcal{A}$ constant and the phase satisfies $\left.\varphi\right|_{\rm K}=D_{\rm KS}$. With no lens, the amplitude receives no correction. Furthermore, using the flat-spacetime metric in Eq.~\eqref{NP}, we find that the only non-vanishing spin coefficients are $\rho=-1/s_{\rm K}$ and $\mu=\rho/2$. The affine parameter at the observer is simply given by the source distance, $s_{\rm K}=D_{\rm KS}$. The background Kirchhoff integral \eqref{wave} therefore becomes
\begin{align}\label{spheq}
\Psi_\text{sph}&=\frac{\omega}{2 \ii\pi}\frac{\e^{\ii\omega D_{\rm K}}}{D_{\rm K}}\bigintsss_{\rm K}\dd^2\zeta \,A_0\left[ 1+\frac{\ii}{\omega}\left( \frac{\Im(A_1)}{A_0}-\frac{\rho}{2}+\frac{1}{2D_{\rm K}} \right) \right. \nn\\
& \left.+\left( \frac{\mathcal{K}_7}{A_0}+\frac{\mathcal{K}_1}{12}-\frac{3}{2}\frac{1}{D_{\rm K}^2} \right)\zeta\bar\zeta +\frac{\ii}{2}\left( \frac{\mathcal{K}_6}{12}-\frac{1}{D_{\rm K}^3} \right)\omega(\zeta\bar\zeta)^2\right] \nn\\
&\hspace{1cm} \times \exp\left[\ii\omega\zeta\bar\zeta\left(\frac{1}{D_{\rm K}}-\rho\right)\right] \e^{\ii\omega D_{\rm KS}} \,.
\end{align}
The $\mathcal{K}$ coefficients are gathered in Appendix~\ref{coeff}, and keeping only the flat background contributions we obtain
\begin{align}
\Psi_\text{sph}&=\frac{\omega}{2 \ii\pi}\frac{\e^{\ii\omega D_{\rm S}}}{D_{\rm K}}\bigintsss_{\rm K}\dd^2\zeta \,\frac{\mathcal{A}}{D_{\rm KS}}\left[ 1+\frac{\ii}{\omega}\left( \frac{1}{2D_{\rm KS}}+\frac{1}{2D_{\rm K}} \right) \right. \nn\\
& \left. -\frac{3}{2}\left( \frac{1}{D_{\rm KS}^2}+\frac{1}{D_{\rm K}^2} \right)\zeta\bar\zeta -\frac{\ii}{2}\left( \frac{1}{D_{\rm KS}^3}+\frac{1}{D_{\rm K}^3}\right)\omega(\zeta\bar\zeta)^2\right] \nn\\
&\hspace{1cm} \times \exp\left[\ii\omega\zeta\bar\zeta\left(\frac{1}{D_{\rm K}}+\frac{1}{D_{\rm KS}}\right)\right] \,.
\end{align}
Performing the Gaussian integrals with \eqref{Gint}, we get
\begin{align}\label{resintsph}
& \Psi_\text{sph}=\frac{\mathcal{A}\e^{\ii\omega D_{\rm S}}}{D_{\rm S}}\Bigg[1 +\frac{\ii}{\omega}\left( \frac{1}{2D_{\rm KS}}+\frac{1}{2D_{\rm K}} \right)- \nn\\
& -\frac{\ii\, L_{\rm K}}{\omega}\frac{3}{2} \left( \frac{1}{D_{\rm KS}^2}+\frac{1}{D_{\rm K}^2} \right) +\frac{\ii\, L_{\rm K}^2}{\omega}\left( \frac{1}{D_{\rm KS}^3}+\frac{1}{D_{\rm K}^3} \right) \Bigg] \,.
\end{align}
Using \eqref{L}, the combination of distances at order $1/\omega$ cancel exactly, and we are left with
\begin{align}\label{sphwave}
\Psi_\text{sph}=\frac{\mathcal{A}\e^{\ii\omega D_{\rm S}}}{D_{\rm S}} \,.
\end{align}
In the absence of a lens, a spherical wave propagating on a flat background remains unchanged. Consequently, no BE corrections arise, as expected, since the spherical wave is an exact solution of the wave equation in flat spacetime.

\subsection{Effect of amplitude perturbations}

We now consider a wave whose amplitude varies across the integration plane, as a simplified model of the BE corrections studied below. We introduce an amplitude perturbation described by $\delta \bar\delta A_0$, which enters explicitly in the wave expression \eqref{wave}. We then evaluate the resulting contribution to the Kirchhoff integral.
This simplified case highlights how transverse variations of the amplitude give rise to BE effects. It serves as an intermediate step toward the complete calculation, while keeping the number of contributing terms manageable.
\begin{figure}[h]
\includegraphics[width=0.7\columnwidth]{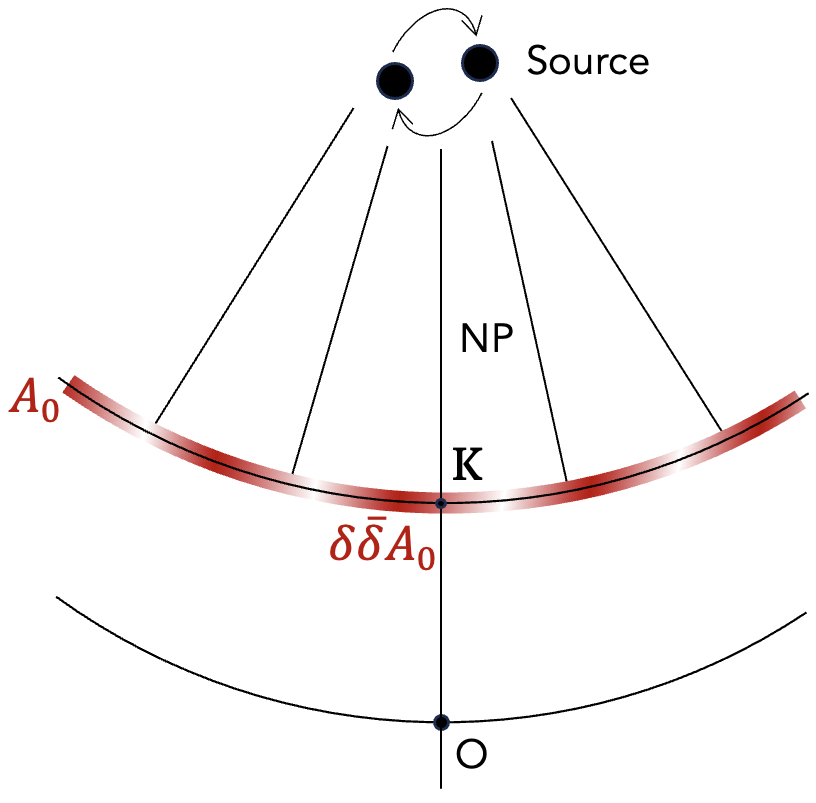}
\caption{Schematic picture representing the system where the amplitude $A_0$ varies on the wave front and gives a non vanishing $\delta\bar\delta A_0$. We perform the Kirchhoff integral on the surface that crosses the geodesic in the point K. To compute the field around K, we evolve the NP scalars from the source position to K.}
\label{ddA0}
\end{figure}
In Eq.~\eqref{wave}, all higher-order quantities vanish except $\delta\bar\delta A_0$, which is non-zero by assumption. The other contributions are therefore the same as those obtained for the spherical wave in Eq.~\eqref{spheq}. The only modification appears in $\mathcal{K}_7$, where the contribution from $\delta\bar\delta A_0$ must be taken into account. Evaluating the integral then gives
\begin{align}\label{resint}
\Psi=&\Psi_\text{sph}\Bigg[1 +\frac{\ii}{\omega}\left( \frac{1}{2D_{\rm KS}}+\frac{1}{2D_{\rm K}} \right) +\frac{\ii\, L_{\rm K}^2}{\omega}\left( \frac{1}{D_{\rm KS}^3}+\frac{1}{D_{\rm K}^3} \right) \nn\\
& +\frac{\ii\, L_{\rm K}}{\omega} \left( \frac{\delta\bar\delta A_0}{A_0}\Bigg\vert_{\rm K}-\frac{3}{2}\frac{1}{D_{\rm KS}^2}-\frac{3}{2}\frac{1}{D_{\rm K}^2} \right) \Bigg] \,.
\end{align}
The combination of distances is the same as for the spherical wave and therefore cancels exactly. The wave then simplifies to
\begin{align}\label{KR1}
\Psi=\Psi_\text{sph}\left[1 +\frac{\ii}{\omega}\frac{\delta\bar\delta A_0\vert_{\rm K}}{\mathcal{A}}s_{\rm K}L_{\rm K} \right] \,.
\end{align}
Here, the term $\Psi_{\rm sph}$ is the spherical wave \eqref{sphwave} that was derived in the previous section.

We can also compute the perturbed wave using the BE formalism summarized in Sec.~\ref{previous}. In this case, the wave at the observer is simply given by
\be\label{KR2}
\Psi=\Psi_\text{sph}\left[ 1+\frac{\ii}{\omega}\frac{\Im(A_1)}{A_0} \right] \,.
\ee
We then use the evolution equation of $A_1$, Eq.~\eqref{evolA1} in our setup. In this case, the only contribution to the source term in Eq.~\eqref{BoxA0} comes from $\delta\bar\delta A_0/A_0$. 
This yields 
\be\label{KR3}
\frac{\Im(A_1)}{A_0} = \frac{\delta\bar\delta A_0\vert_{\rm K}}{\mathcal{A}}s_{\rm K}L_{\rm K} \,.
\ee
We have therefore shown that the Kirchhoff integral result~\eqref{KR1} and the BE formalism, that is~\eqref{KR2} with~\eqref{KR3}, are equivalent. This preliminary exercise validates the approach and prepares the ground for the full computation in the presence of a matter lens, presented in the following section.

\subsection{General computation at linear order} \label{linear}

Finally, we evaluate the improved diffraction integral \eqref{wave}  for a matter lens at linear order in the gravitational potential. The integral is computed separately for each image, and the complete result is obtained by summing over all images. 
As we shall find that the corrections vanish when the beam remains in pure vacuum, we also derive in Appendix~\ref{PointLikeLens} the corresponding expression for Weyl lensing (e.g. for a point-like lens), restricting to the leading quadratic contribution in the potential.

We focus on a single GO trajectory. Let K denote the point on the geodesic about which the Kirchhoff integral is evaluated, with corresponding affine parameter $s_{\rm K}$. We choose K to lie beyond the lens, such that $s_{\rm L}\leq s_{\rm K}\leq s$. The point K therefore specifies the location of the lens plane on which the Kirchhoff integral is performed. 
The coefficients $C_n$ in the expansion~\eqref{gaussian} are decomposed as
\be
\mathcal{C}_n = \mathcal{C}_n^{\rm GO} + \frac{1}{\omega}\left[\mathcal{C}_n^{\rm BE}+\mathcal{C}_n^{\rm sph}\right]\,,
\ee
that is we classify each contribution according to its order in the high-frequency expansion: whether it contributes to GO, i.e., at zeroth order in $\omega$, or whether it gives rise to a contribution of order $1/\omega$. In the latter case, one can further distinguish between the spherical contributions, which combine to vanish at each order in $1/\omega$ and do not depend on the lens, and the BE terms, which represent corrections to the geometric-optics results arising from wave effects induced by the lens. 
We also decompose the NP scalars into their zeroth and first order contributions in $\phi$. In particular, we write the expansion rate as $\rho=\rho^{(0)}+\rho^{(1)}$ and $\mu=\mu^{(0)}+\mu^{(1)}$. Since we expand the exponential around its flat-spacetime value, for which only $\rho^{(0)} $ enters the Hessian, the first-order contribution $\rho^{(1)}$ is expanded out of the exponential and must therefore be explicitly kept in the integrand. By contrast, all other spin coefficients vanish at zeroth order in the lens mass, so we omit the superscript (1), with the understanding that they are first-order quantities. Finally, we emphasize that each spin coefficient is obtained by integrating its propagation equation along the geodesic. Consequently, its projected derivatives constitute independent quantities, which must likewise be determined by solving their corresponding propagation equations. The evolution equations for each NP scalar can be found in \cite{Bruyere:2026gnt}. Their solutions are also provided in Appendix~\ref{NPsol}.

The coefficient $C_n$ are obtained by recasting~\eqref{wave} under the form~\eqref{gaussian}. From section \ref{sph}, we find the geometric-optics terms 
\be \label{GO}
\mathcal{C}_0^{\rm GO}= 1 \,,\quad  \mathcal{C}_1^{\rm GO} = -\ii \rho^{(1)}\,,\quad \mathcal{C}_2^{\rm GO}=\mathcal{C}_3^{\rm GO}=0\,.
\ee
The "spherical" terms are
\begin{align}
&\mathcal{C}_0^{\rm sph}= \frac{\ii}{2}\left(\frac{1}{(s-s_{\rm K})}-\rho^{(0)}\right) ,\\
&\mathcal{C}_1^{\rm sph}=-\frac{3}{2}\left(\frac{1}{(s-s_{\rm K})^2}+(\rho^{(0)})^2\right) ,\\
&\mathcal{C}_2^{\rm sph}=-\frac{\ii}{2}\left(\frac{1}{(s-s_{\rm K})^3}-(\rho^{(0)})^3\right)\,,
\end{align}
and the BE contributions appearing in Eq.~\eqref{wave} are
\begin{align}
&\mathcal{C}^{\rm BE}_0 = \frac{A_1}{A_0} -\frac{\ii}{2}\rho^{(1)}\,, \\
&\mathcal{C}^{\rm BE}_1 = \frac{\delta \bar \delta A_0}{A_0} + \frac{1}{3}\Re(\delta \bar \tau)+\rho^{(1)}\left(\frac{1}{2(s-s_{\rm K})}-\frac{17}{6}\rho^{(0)}\right) \nn\\
&\hspace{1,2cm} -\frac{4}{3}\rho^{(0)}\mu^{(1)} \,,\\
&\mathcal{C}_2^{\rm BE} = \ii\left[\rho^{(1)}\left(\frac{31}{12}(\rho^{(0)})^2+\frac{3}{2(s-s_{\rm K})^2}\right)+\frac{5}{6}(\rho^{(0)})^2\mu^{(1)}\right. \nn\\
&\hspace{0,5cm} \left.-\frac{1}{6}\delta \bar \delta \rho - \frac{1}{12}\Re(\delta \delta\bar\sigma +4\bar\sigma\delta\beta)-\frac{2}{3}\rho^{(0)}\Re(\delta \bar \tau)\right], \label{C2BE}\\
&\mathcal{C}_3^{\rm BE} = \frac{1}{2}\rho^{(1)}\left((\rho^{(0)})^3-\frac{1}{(s-s_{\rm K})^3}\right).
\end{align}
The Gaussian integral \eqref{gaussian} is evaluated using the identities in \eqref{Gint}. Since the correction $A_1$ is purely imaginary (after choosing the geometric-optics amplitude $A_0$ to be real, as in \cite{Bruyere:2026gnt}), we may take the imaginary part of Eq.~\eqref{gaussian} to isolate the correction proportional to $\Im(A_1)$. We find
\begin{align} \label{AKirch}
\omega\frac{\Im(A_1)}{A_0} &=\omega\frac{\Im(A_1)}{A_0}\Big\vert_{\rm K} + L_{\rm K}\left(\frac{\delta \bar \delta A_0}{A_0} + \frac{1}{3}\Re(\delta \bar \tau)\right)\Big\vert_{\rm K} \nn\\
&+\frac{L_{\rm K}^2}{3}\left(\delta \bar \delta \rho + \frac{1}{2}\Re(\delta \delta \bar \sigma+ 4 \delta \beta \bar \sigma)-4 u_{\rm K} \Re(\delta \bar \tau)\right)\Big\vert_{\rm K} \nn\\
&+ \frac{L_{\rm K}u_{\rm K}}{3}\left(\frac{5}{2}L_{\rm K}u_{\rm K} -2\right)\left(\rho^{(1)}-2\mu^{(1)}\right)\Big\vert_{\rm K}\,.
\end{align}
The NP scalars on the r.h.s are evaluated at $u_{\rm K}$, and with expressions in Appendix~\ref{NPsol}, they are given by 
\begin{flalign} 
& \left( \rho^{(1)}-2\mu^{(1)} \right)\Big\vert_{\rm K} = 4\pi G \Sigma\frac{u_{\rm K}}{u_{\rm L}} \,, \label{NPatD1} &&\\ 
& \left( \frac{\delta \bar \delta A_0}{A_0} + \frac{1}{3}\Re(\delta \bar \tau) \right)\Big\vert_{\rm K} = 4\pi G \Sigma\frac{u_{\rm K}^2}{u_{\rm L}}\left(\frac{2}{3}- \frac{u_{\rm K}}{u_{\rm L}} \right) \label{NPatD2} &&\\
&\hspace{3,8cm} +2\pi G \nabla^2_{\perp} \Sigma \frac{u_{\rm K}^2}{u_{\rm L}^3} \left(1-\frac{u_{\rm K}}{u_{\rm L}}\right) , &&\nn\\ 
& \left( \frac{1}{3}\delta \bar \delta \rho + \frac{1}{6}\Re(\delta \delta \bar \sigma+ 4 \delta \beta \bar \sigma)-\frac{4}{3}u_{\rm K} \Re(\delta \bar \tau) \right)\Big\vert_{\rm K} \label{NPatD3} &&\\
&\hspace{2cm} = 2\pi G \Sigma\frac{u_{\rm K}^3}{u_{\rm L}}\left(\frac{u_{\rm K}}{u_{\rm L}} -\frac{5}{3}\right) + \pi G \nabla^2_{\perp} \Sigma \frac{u_{\rm K}^4}{u_{\rm L}^4} \,. &&\nn 
\end{flalign}
Substituting these expressions and \eqref{A1A0u} into Eq.~\eqref{AKirch}, we find that the field at the observer can be expressed, to linear order in $\phi$, as 
\be\label{IJ}
\Psi^{(j)}=\Psi_{\text{GO}}^{(j),\mu=1}\left(1+\Delta^{\rm mag}_j+\frac{\ii}{\omega}\Delta_j\right) \,,
\ee
where the GO part $\Psi_{\text{GO}}^{(j),\mu=1}$ appears in Eq.~\eqref{Psimag} and the correction $\Delta^{\rm mag}_j$ in \eqref{Dmag}. We check that the result does not depend on the choice of integration point after the lens. Therefore, the integration can equivalently be carried out at the lens plane, which simplifies the expressions~\eqref{NPatD1}-\eqref{NPatD3} by setting $u_{\rm K}=u_{\rm L}$.
The resulting BE correction, which can be interpreted as a phase correction, is
\begin{align} 
\Delta_j&=\frac{\Im\left(A_1\right)}{A_0}\nn\\
&= \pi G\nabla^2_{\perp}\Sigma\left(\frac{u_{\rm L}-u}{u_{\rm L}^2}\right)^2+2\pi G \Sigma  \left(\frac{u}{u_{\rm L}}\right)^2 \label{Delta_j}\\
&= \pi G\nabla^2_{\perp} \Sigma \left( \frac{D_{\rm L} D_{\rm LS}}{D_{\rm S}}\right)^2 +2\pi G \Sigma  \left(\frac{D_{\rm LS}}{D_{\rm S}}\right)^2 \label{importantresult1}\,.
\end{align}
Note that all distances appearing on the right-hand side are evaluated for the $j$-th image. The result~\eqref{importantresult1} coincides with Eq.~(107) of \cite{Bruyere:2026gnt}, also summarized in Sec.~\ref{previous}, where it was obtained by solving the BE system of equations from the source to the observer. The equivalence between the two approaches is therefore established, as expected.
The correction $\Delta_j$ can be decomposed as in Eq.~\eqref{2Delta} into two contributions: a {\it{standard}} term, which is already captured by the standard derivation, and a {\it{new}} term, corresponding to an additional contribution absent from the standard treatment. The total field is eventually obtained by summing over the contributions from all images.

\subsection{Non-linear contributions}\label{NL}

Generalizing the previous results to higher orders in $\phi$ is a formidable task. However, we expect that the dominant non-linear effects are those which, once resumed, combine to give the magnification prefactor in~\eqref{Fst}. They arise from the modification of the quadratic part of the phase in the exponential due to lensing. Therefore in order to capture a similar effect in the Kirchhoff integral we must keep the exponential in~\eqref{wave} and not expand it around the flat space expression as was performed in the decomposition~\eqref{gaussian}.

Even this simplified treatment of non-linearities is a very complicated task, hence we also restrict to an axisymmetric beam, such that we can assume that $\sigma = 0$ in the quadratic part of the exponential. With these assumptions, the magnification is only due to $\rho$ departing from the flat spacetime solution thanks to matter convergence, and the  magnification reduces to
\be
\sqrt{\mu} = \frac{1}{1-4\pi G \Sigma D_{\rm L}D_{\rm LS}/D_{\rm S}}\,.
\ee
In this setting, either the image has not crossed any caustic and $\sqrt{\mu} >0$ or it is fully inverted and $\sqrt{\mu}<0$.

In practice in eq.~\eqref{wave} we will encounter integrals of the form
\be\label{Gint2}
\int \dd^2\boldsymbol{x} \,{g}(\boldsymbol{x}) (\omega\zeta \bar \zeta)^n = n! \sqrt{\mu} (\ii L_{\rm K} \sqrt{\mu})^n\,,
\ee
where
\be
g(\boldsymbol{x}) \equiv \frac{\omega}{2\pi \ii L_{\rm K}}\exp\left(\ii \omega \frac{2\zeta\bar\zeta}{2L_{\rm K} \sqrt{\mu}}\right)\,.
\ee
Eventually we find
\be\label{IJNL}
\Psi^{(j)}=\Psi_{\text{GO}}^{(j)}\left(1+\frac{\ii}{\omega}\Delta^{\rm NL}_j\right) \,,
\ee
where $\Psi_{\text{GO}}^{(j)}$ is \eqref{psiGOj} and
\begin{align}\label{nonlinearresult}
\Delta^{\rm NL}_j =&\, \pi G\nabla^2_{\perp} \Sigma \left( \frac{D_{\rm L} D_{\rm LS}}{D_{\rm S}}\right)^2 \sqrt{\mu}^2\\
&+2\pi G \Sigma  \sqrt{\mu}\frac{D_{\rm L}\left(4 D_{\rm S} - 5 D_{\rm L}\sqrt{\mu}\right)}{D_{\rm S}^2}\nonumber\\
&+(\sqrt{\mu}-1)\frac{\left[-D_{\rm S}^2 + 2 \sqrt{\mu}(D_{\rm L}^2 - D_{\rm L}D_{\rm LS}+D_{\rm LS}^2)\right]}{2 D_{\rm L}D_{\rm S} D_{\rm LS}}\nonumber\,.
\end{align}
First, it can be checked that we recover the first order result~\eqref{importantresult1} when expanding $\sqrt{\mu}$ at first order in $G \Sigma$. For large magnifications we note that the subdominant contribution (everything but the first line) scales as $\sqrt{\mu}^2/D_{\rm LS}$ when $D_{\rm L} \gg D_{\rm LS}$ hence it does not fade away. However, it is suppressed by a factor of order $b^2/D_{\rm LS}^2$ with respect to the dominant contribution (the first line) which is also enhanced by an extra $\sqrt{\mu}^2$. 

We stress that even though~\eqref{nonlinearresult} is more general than~\eqref{importantresult1}, it is not valid in all situations given our simplified treatment of non-linearities. In particular if lensing is dominated by the Weyl contribution, as is the case if the beam crosses only an empty region around the lens, our assumption of beam axisymmetry is not valid. The dominant contribution at order $G^2$ in that case is derived in Appendix \ref{PointLikeLens} where the role of the shear is crucial.

\subsection{Mapping to standard Kirchhoff evaluation}\label{mapping}

This section clarifies the nomenclature "standard" and "new" used for the two contributions in the phase correction at linear order \eqref{2Delta}.  We compare our result with the standard evaluation of the Kirchhoff integral (the diffraction integral), given by Eq.~\eqref{finalSt}, and identify the terms that are neglected in the latter approach.

First, it is easy to check that the quadratic parts of the phase expansion are equivalent. The Hessian matrix of the diffraction integral is~\eqref{DefHab}. In the improved integration it is given by the matrix in the r.h.s. of~\eqref{defHbarab2} that we must evaluate at the lens plane ($s_K = s_L$). At first order $\rho$ and $\sigma$ are given for a thin lens by~\eqref{rhofirstorder} and~\eqref{sigmafirstorder}, and using $(\partial_{xx}+\partial_{yy})\psi = 4\pi G \Sigma$ it reduces to~\eqref{DefHab} with $\bar{L} \to L$, a small difference already found at the end of section~\ref{eikonal_evolution}.

We can therefore focus on the higher-order contributions, and specifically the quartic term. We compare the standard quartic contribution in Eq.~\eqref{Tau} with the quartic terms extracted from the NP scalars in the BE formalism \eqref{wave}, still at linear order in $\phi$. 
To perform this comparison, we choose the same coordinate system in both approaches, namely a plane orthogonal to the geometric-optics trajectory rather than to the optical axis. This plane can be described either by the polarization basis ($\boldsymbol{m},\boldsymbol{\bar m}$) or the flat basis ($\boldsymbol{e}_x,\boldsymbol{e}_y$). 
We stress once again that, since the choice of the integration surface is arbitrary, this procedure yields the same result as the expansion performed around the lens plane in Sec.~\ref{StF}. However, the present choice is more convenient for matching the two approaches term by term.

As in \eqref{tdex}, we expand the time delay around a given image $j$.
Since the geometric contribution to the time delay is quadratic in the transverse distances [Eq.~\eqref{Tgeom}], any derivative of order higher than two acting on this term vanishes. Consequently, the quartic contribution only receives contributions from the gravitational part of the time delay. Expanding the integrand of Eq.~\eqref{finalSt} as in Eq.~\eqref{FFF}, we find that the quartic BE correction at linear order in $\phi$ is 
\begin{align} \label{ddddT}
& x^ax^bx^cx^d \nabla_a\nabla_b\nabla_c\nabla_d \,T \supset m^am^b\bar m^c\bar m^d \nabla_a\nabla_b\nabla_c\nabla_d \,T \nn \\
&\hspace{4cm} \times 6(\zeta\bar\zeta)^2 \,.
\end{align}
The symbol $\supset$ indicates that we keep only the terms that give a non-vanishing contribution after performing the Gaussian integral.
Using this result, we can rewrite Eq.~\eqref{finalSt} in the form
\begin{align}\label{Psicorr}
\Psi^{\text{st}}&=\frac{\omega A }{2\pi \ii D_{\rm L}} \int d^2\bm{\zeta}\, e^{\ii \omega T_{\rm GO}} \nn\\ 
&\times\left[ 1 + \frac{\ii}{4} \omega(\zeta\bar\zeta)^2 m^am^b\bar m^c \bar m^d\nabla_a\nabla_b\nabla_c\nabla_d \,T \right] .
\end{align}
We now identify which term in Eq.~\eqref{wave} gives rise to this first BE correction. 
Since it involves four derivatives of the potential $\phi$, we need to identify the corresponding terms in the full integral \eqref{wave} that can reproduce this structure.
The expression in Eq.~\eqref{wave} is written in terms of NP scalars, whose dependence on the lens potential can be determined from their evolution equations. In these equations, the Weyl and Ricci scalars are the quantities that depend explicitly on $\phi$. Their expressions at linear order in $\phi$ can be obtained following the procedure described in Sec.~2.11 of \cite{Bruyere:2026gnt}. Using the evolution equations, we find that the only NP combinations containing four derivatives of the potential are $\delta\bar\delta\rho$ and $\delta\delta\bar\sigma+4\bar\sigma\delta\beta$ whose evolution is dictated by
\begin{align}
&D\delta\bar\delta\rho=4\rho\delta\bar\delta\rho+\delta\bar\delta\Phi_{00} +\mathcal{O}(\phi^2) \,,\label{evddrho} \\
&D\left(\delta\delta\bar\sigma + 4\bar\sigma\delta\beta \right) =4\rho\left( \delta\delta\bar\sigma + 4\bar\sigma\delta\beta \right) +\delta \delta \bar{\Psi}_0 \nn\\
&\hspace{2,8cm} +8 \delta \beta \bar \delta \bar \delta \phi +\mathcal{O}(\phi^2) \label{evddsig}\,.
\end{align}
These terms appear in $\mathcal{C}_2^{\rm BE}$, Eq.~\eqref{C2BE}. Indeed, they scale as $\omega(\zeta\bar\zeta)^2$, see Eq.~\eqref{gaussian}, and therefore have the same scaling as the standard correction in Eq.~\eqref{Psicorr}.
From the weak-field analysis of \cite{Bruyere:2026gnt}, we have $\delta\bar\delta\Phi_{00}=\delta\bar\delta\nabla^2\phi$ which contains the term $2\delta\bar\delta\delta\bar\delta\phi$, and $\delta\delta\bar\Psi_0=2\delta\delta\bar\delta\bar\delta\phi$. Combining these contributions, we extract the structure $2\,m^am^a\bar m^c \bar m^d \nabla_a \nabla_b \nabla_c \nabla_d \,\phi$. Integrating the corresponding evolution equations, \eqref{evddrho} and \eqref{evddsig}, and using Eq.~\eqref{TGrav}, we can rewrite the result in terms of the gravitational time delay as $m^am^a\bar m^c \bar m^d \nabla_a \nabla_b \nabla_c \nabla_d \,T$. Collecting all the coefficients, we find the combination appearing in $\mathcal{C}_2^{\rm BE}$ which reproduces exactly the BE correction obtained from the diffraction integral in Eq.~\eqref{Psicorr}, and it is
\begin{align}
\mathcal{C}_2^{\rm BE} &\supset -\frac{\ii}{6}\delta \bar \delta \rho -\frac{\ii}{12}\Re(\delta \delta\bar\sigma +4\bar\sigma\delta\beta) \\
& \supset \frac{\ii}{4}m^am^b\bar m^c \bar m^d\nabla_a\nabla_b\nabla_c\nabla_d \,T \nn\,.
\end{align}
Therefore, we can pinpoint that the contributions involving four derivatives of the lens potential are precisely the ones that are taken into account by the standard diffraction integral, and they lead to $\Delta^{\rm st}$ exclusively. Consequently, the other contributions are responsible for the appearance of $\Delta^{\rm new}$.

\section{Application: singular isothermal sphere}\label{SIS}

\subsection{Geometry of the singular isothermal sphere}

We now evaluate the corrections (\ref{Delta_j}) for a realistic lens model. We consider a singular isothermal sphere (SIS) \cite{1992grle.book}, which provides a simple yet realistic description of a galaxy acting as a gravitational lens. For illustrative purposes, we restrict the analysis to linear order in the gravitational potential $\phi$.
The matter density profile of a singular isothermal sphere is given by
\be\label{rhoSIS}
\rho_{\rm mat}(r)=\frac{\sigma_v^2}{2\pi G r^2}\,,
\ee
where $\sigma^2_v$ is the velocity dispersion of the lens. Despite its central singularity and infinite total mass, the SIS provides a reasonably realistic model for the mass distribution of a galaxy acting as a gravitational lens. We assume the velocity dispersion to be constant throughout the lens.

The metric for a static, spherically symmetric spacetime is
\be\label{metricSIS}
\dd s^2 = -\e^{2\Phi(r)}\dd t^2+f(r)^{-1}\dd r^2+r^2\dd^2\Omega \,,
\ee
with $f(r)=(1-2Gm(r)/r)$ and $m(r)$ is the mass function, while $\Phi$ is the gravitational potential, which in the weak-field limit is related to the Newtonian potential via
\be
e^{2\Phi}=1+2\phi \,.
\ee
The Tolman–Oppenheimer–Volkoff (TOV) equations~\cite{Tolman1939,OppenheimerVolkoff1939}, obtained from Einstein equations together with the condition of hydrodynamic equilibrium, govern the structure of a spherical lens and are given by
\begin{align}\label{TOVeq1}
&\frac{{\rm{d}} p}{{\rm{d}} r}=-(p+\rho_{\rm mat})\frac{\rm{d}\Phi}{{\rm{d}}r} \,,\,\,\,\, \frac{{\rm{d}}m}{{\rm{d}}r}=4\pi\rho_{\rm mat} r^2 \,, \\
&\frac{{\rm{d}}\Phi}{{\rm{d}}r}=G\frac{m+4\pi r^3 p}{r(r-2Gm)} \,,\label{TOVeq2}
\end{align}
where $\rho_{\rm mat}(r)$ is the matter density and $p(r)$ is the matter pressure. Substituting the SIS density profile \eqref{rhoSIS} into Eqs.~\eqref{TOVeq1} allows us to derive the explicit forms of the mass and pressure profiles for a SIS lens
\be
m(r)=\frac{2\sigma_v^2}{G}r \quad\text{and}\quad p(r)=\sigma_v^2\,\rho_{\rm mat}(r) \,.
\ee
Then, using these results in (\ref{TOVeq2}) we get  the radial profile of the gravitational potential 
\be\label{PhiSIS}
\Phi(r)= 2\sigma_v^2\alpha\ln\left(\frac{r}{r_0}\right)\,,
\ee
where $\alpha=(1+\sigma_v^2)/(1-4\sigma_v^2)$.
Considering a geodesic with impact parameter $b$, the null geodesic condition gives
\be
\frac{\dd r}{\dd s}=\mp \sqrt{f(r)}\e^{-\Phi(r)}\sqrt{ 1-\frac{b^2}{r^2}\e^{2\Phi(r)} } \,.
\ee
We observe that in the weak-field limit, and using isotropic coordinates, the metric (\ref{metricSIS}) can be written as
\be
\dd s^2 = -\left(1+2\phi(r)\right)\dd t^2+\left(1-2\phi(r)\right) \left(\dd r^2+r^2\dd^2\Omega\right) \,,
\ee
where the Newtonian lens potential $\phi$ is given by \eqref{PhiSIS} at linear order in $\sigma_v^2$. 
For the SIS density profile, we can compute the surface density $\Sigma$ and $\nabla^2_{\perp} \Sigma$, and hence get $\Im(A_1)/A_0$. At linear order in the potential, they are given by 
\begin{align}
& \Sigma=\frac{\sigma_v^2}{2Gb} \,, \\
& \nabla^2_{\perp} \Sigma=\frac{\sigma_v^2}{Gb^3} \,,
\end{align}
where $b=D_{\rm L}\sin(\theta)$ is the impact parameter corresponding to each image position $\theta_j$. It is determined by the source position $\beta$ through the lens equation. A schematic representation of the lensing geometry is shown in Fig.~\ref{setup}.

\subsection{Lensing equation}

We now introduce the standard notation used in strong lensing. We denote $x$ and $y$ as the rescaled angular positions of the image and the source, respectively (not to be confused with the coordinates $x$ and $y$ used earlier in this article). 
More explicitly, we define $\bx_j=\boldsymbol{\theta}_j/\theta_E$ and $\mathbf{y}=\boldsymbol{\beta}/\theta_E$, where $\boldsymbol\theta_j$ is the angular position of the $j$-th image, $\boldsymbol\beta$ is the source angular position of the source, and $\theta_E$ is an arbitrary angular scale, chosen here to be the Einstein radius, defined as
\be
\theta_E=4\pi\sigma_v^2\frac{D_{\rm LS}}{D_{\rm S}}\,.
\ee
The lens equation can be written as
\be
\mathbf{y}=\mathbf{x}-\frac{\mathbf{x}}{|\mathbf{x}|} \,,
\ee
and for $y<1$ it has two solutions 
\be
|\mathbf{x}_1|=x_1=1+y\,,\qquad |\mathbf{x}_2|=x_2=1-y\,,
\ee
where $y=|\mathbf{y}|$. The two solutions correspond to images of opposite parity. For $y=0$, the solution reduces to the Einstein ring, characterized by $|\mathbf{x}|=1$. For $y>1$, only one image remains, as the second solution no longer exists.

\subsection{Time delay}

Let us consider the solutions of the lens equation and derive the frequency-dependent time delay between the two images, including both the geometric-optics contribution and the diffraction-induced corrections. The GO time delay between images 1 and 2 is given by (see e.g. \cite{1992grle.book}) 
\be
\Delta T_{\rm GO}=\theta_E^2\frac{D_{\rm L}D_{\rm S}}{D_{\rm LS}}2y=(4\pi\sigma_v^2)^2\frac{D_{\rm LS}D_{\rm L}}{D_{\rm S}}2y \,,
\ee
where $2y =x_1-x_2$.
When diffraction effects are taken into account, the phase of the wave associated with each GO image receives a correction $\varphi\rightarrow\varphi+\omega^{-2}\Delta_j$, with $\Delta_j$ given by Eq.~\eqref{Delta_j}. For a SIS lens, this correction becomes, in the small-angle approximation,
\begin{align}\label{A1sis}
\Delta_j &= \frac{\pi\sigma_v^2}{x_j\theta_ED_{\rm L}} \left( \frac{D_{\rm LS}}{D_{\rm S}}\right)^2 + \frac{\pi\sigma_v^2}{(x_j\theta_ED_{\rm L})^3} \left( \frac{D_{\rm LS}D_{\rm L}}{D_{\rm S}}\right)^2\,,
\end{align}
In Eq.~\eqref{A1sis}, the second contribution corresponds to the term already present in the standard diffraction approach, while the first one represents the new contribution derived in this work. The standard term is enhanced relative to the new contribution by a factor $(D_L/b)^2$. 
Therefore, BE corrections induce a modification of the time delay difference between the two images, given by
\begin{align}
\Delta T_{\rm BE} &= \Delta T_{\rm st}+\Delta T_{\rm new}\,,
\end{align}
where 
\begin{align}
&\Delta T_{\rm st}= \omega^{-2}\frac{\pi\sigma_v^2}{(D_{\rm L}\theta_E)^3}\left( \frac{D_{\rm LS}D_{\rm L}}{D_{\rm S}}\right)^2\left[ \frac{1}{(1+y)^3} - \frac{1}{(1-y)^3} \right] , \\
&\Delta T_{\rm new}= \omega^{-2}\frac{\pi\sigma_v^2}{D_{\rm L}\theta_E}\left( \frac{D_{\rm LS}}{D_{\rm S}}\right)^2\left[ \frac{1}{(1+y)} - \frac{1}{(1-y)} \right] .
\end{align}
\begin{figure}[!ht]
\centering
\includegraphics[width=1\columnwidth]{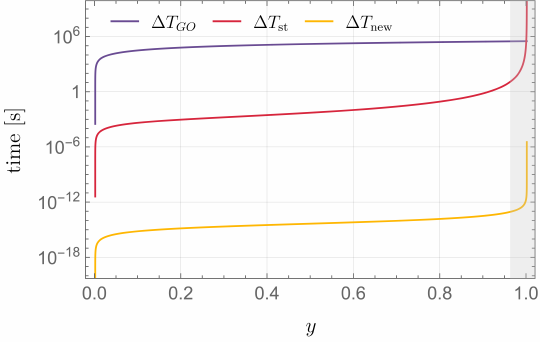}
\caption{Time delay between the two images in GO and with the diffraction effects from the standard diffraction term, $\Delta T_{\rm st}$, and the new one, $\Delta T_{\rm new}$. We have chosen system with a frequency of $10^{-3}\,{\rm Hz}$, $D_{\rm LS}=100\,{\rm Mpc}$, $D_{\rm L}=150\,{\rm Mpc}$ and $\sigma_v=200\,{\rm km/s}$.}
\label{im0}
\end{figure}
We observe a divergence of $\Delta T_{\rm st}$ and $\Delta T_{\rm new}$ at $y=1$, which originates from the singularity of the density profile at the centre of the SIS. Indeed for $y=1$, one of the images crosses the centre of the lens distribution.

We stress that our approach is valid provided that
\begin{equation}\label{validity}
\omega>\omega_{\rm ref},,\qquad
\omega_{\rm ref}=\max\{\Delta_1,\Delta_2\},,
\end{equation}
where $\Delta_1$ and $\Delta_2$ denote the phase corrections associated with the two images, as given in Eq.~\eqref{A1sis}. This condition guarantees the validity of the perturbative BE expansion used to compute the field on the lens plane.

Results for a SIS lens model are shown in Fig.~\ref{im0} as function of the source position $y$. We find that BE corrections remain subdominant compared to the leading GO contribution, with the standard diffraction contribution providing the dominant BE correction. The grey region in Fig.~\ref{im0} corresponds to the region of parameter space where the validity condition in Eq.~\eqref{validity} is not satisfied.

\begin{figure*}[!ht]
\centering
\includegraphics[width=1\columnwidth]{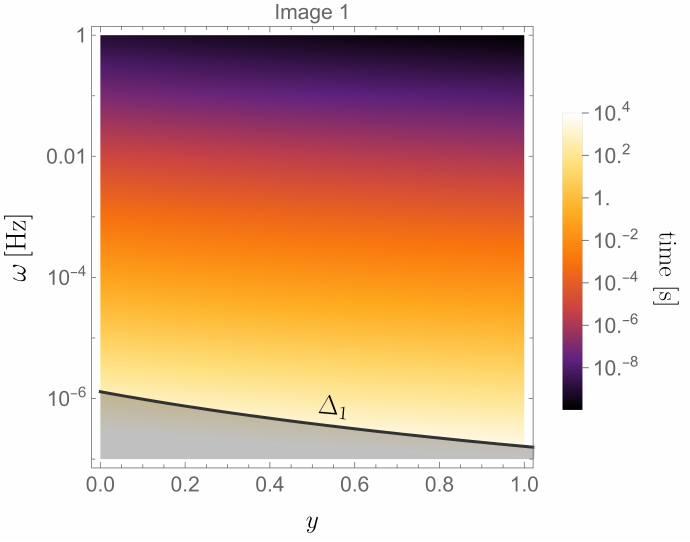}
\includegraphics[width=1\columnwidth]{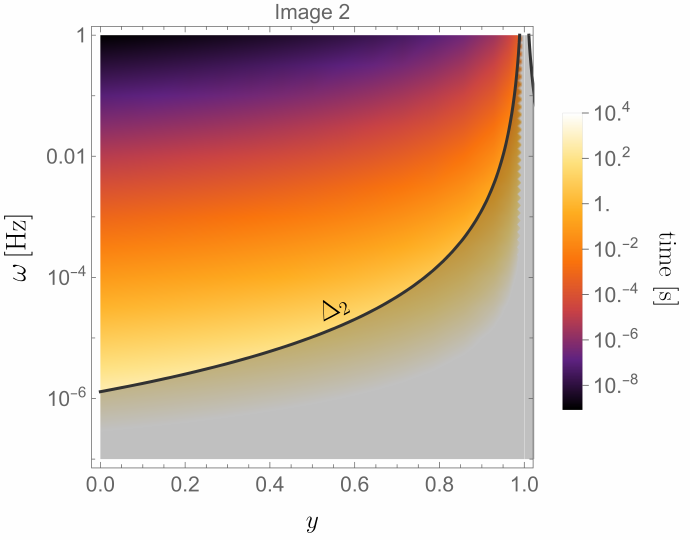}
\caption{\small Total BE time delay for the image 1 and 2 as a function of the source position $y$ and the frequency $\omega$. The black line is frequency limit for the two images and the grey region below is the space where the formalism is not valid.}
\label{im}
\end{figure*}

In Fig.\,\ref{im}, we show the total BE correction to the time delay for each image as a function of the source position $y$ and the frequency $\omega$. The BE contribution becomes increasingly significant in the deep wave regime. However, the wavelength expansion is valid only above the limiting frequency defined by Eq.~\eqref{validity}, which restricts the range over which the wave regime can be explored perturbatively. For the chosen geometry, the limiting frequency is set by the second image. As $y$ approaches unity, the second image moves progressively closer to the lens centre, enhancing the corresponding correction $\Delta_2$. This explains the behaviour of $\Delta_2$; the white band in the right panel corresponds to the region where this expression diverges.

\section{Discussion and conclusions}\label{discussion}

In this work, we have critically revisited the derivation of the diffraction integral, understood as an approximate evaluation of the Kirchhoff integral, which is widely used in the literature for gravitational lensing applications. We have developed a new evaluation of the Kirchhoff integral in the BE regime, that can keep track of all approximations introduced.

Our approach shows that the standard diffraction integral arises as the leading-order term of a controlled expansion, while additional corrections naturally appear when previously neglected effects, such as amplitude variations and aberration, are consistently included. We find that these corrections are typically small in realistic lensing configurations, thereby providing a posteriori justification for the robustness of the diffraction integral in phenomenological studies beyond the GO limit.

A central result of this work is the equivalence between the improved Kirchhoff evaluation and the BE formalism developed in \cite{Bruyere:2026gnt}. In particular, we have shown that propagating the field BE and summing over all geometric-optics images reproduces exactly the same corrections obtained from the systematic expansion of the Kirchhoff integral, hence establishing a direct bridge between both approaches.

This equivalence provides a clearer physical interpretation of diffraction in gravitational lensing. In the BE picture, diffraction arises from contributions in a neighbourhood of geometric-optics trajectories with characteristic Fresnel size. In the Kirchhoff formalism, the same physics emerges from the structure of the integral near the geometric-optics image points once sub-leading corrections are included. We therefore conclude that diffraction effects and BE corrections are not separate physical phenomena, but rather different manifestations of the same underlying wave-optics physics.

As an illustration, we presented a simple SIS lens model and computed the BE corrections to the time delay between the two images, showing that the corrections to the standard diffraction result are very small in most of the parameter space.

Finally, the starting point of this discussion was the equation governing the dynamics of a scalar field. However, the most interesting application is GW diffraction, and a GW is a spin-2 object. Including polarization structure in the description of diffraction is a nontrivial task. Attempts to address this include covariant perturbation theory beyond ray optics \cite{Cusin:2019rmt, Dalang:2021qhu, Harte:2018wni, Dolan:2018ydp}, although the physical interpretation of observables is not straightforward in this context, and path-integral techniques \cite{Braga:2024pik}, whose phenomenological implications have yet to be explored.\footnote{See also \cite{Cusin:2018avf} where a Boltzmann-like approach is used to describe polarization generation via lensing diffusion in a stochastic background context.} Another approach is to exploit an analogy with scattering theory:  phenomenological results have so far been obtained only for very long wavelengths \cite{Pijnenburg:2024btj}. An extension of the present study to spin-2 waves would allow us to map perturbative techniques to diffraction-like treatments and clarify which wave-effects are observable for spin-2 fields. 

\subsection*{Acknowledgements}

We are very grateful to Jean-Philippe Uzan, Julien Larena, Pierre Fleury and Martin Pijnenburg for discussions and feedback. The work of GC is supported by CNRS and SNSF Ambizione grant  PZ00P2-193292.

\appendix

\section{Derivation of the Kirchhoff integral}\label{Kirchhoff}

We consider a scalar field $\hat{\Psi}$ on a spacetime described by the following metric 
\be
\dd s^2=-(1+2 \phi) \dd t^2 +(1- 2 \phi) \dd\bm{x}^2\,,
\ee
where $\phi$ is the spherical gravitational  potential of the lens. The linearized scalar field equation is 
\be
\Box \hat{\Psi}=0\,\quad \Rightarrow \quad  \left(\Delta+\omega^2\right)\Psi= 4\omega^2 \phi \Psi\,,
\ee
with $\hat{\Psi}$ and $\Psi$ related by~\eqref{TotPsivsPsi}. 

We also assume that the thin-lens approximation is valid for the gravitational potential of the lens. We define a volume $V$, as shown in Fig.~\ref{Lens0}, enclosed by a surface close to the lens plane (but not including the lens plane) and containing the observer. Since we assume the thin-lens approximation, $\phi=0$ throughout $V$. We stress that the choice of this surface is arbitrary, provided that the criteria discussed above are satisfied. 
 Then the field equation reduces to 
\be\label{noU}
\left(\Delta+\omega^2\right)\Psi=0\,.
\ee
We can consider a system of spherical coordinates centered on the lens.  The Green's function associated with \eqref{noU} has the form 
\be
\Psi_G=-\frac{1}{4\pi}\frac{e^{\ii\omega r}}{r}\,,\quad \left(\Delta+\omega^2\right)\Psi_G=\delta_{\rm D}^3(\bm x)\,.
\ee
Then one has
\begin{align}
&\Psi_G \Delta \Psi+\omega^2 \Psi_G \Psi=0\,,\\
&\Psi \Delta \Psi_G+\omega^2 \Psi_G \Psi=\Psi \delta_{\rm D}^3(\bm{x}) \,,
\end{align}
hence we get that the field at the observer can be written in terms of the following volume integral 
\begin{align}
\Psi(0)&=\int_V \dd^3\bm{x}\,\left(\Psi\Delta \Psi_G-\Psi_G\Delta \Psi\right)\nn\\
&= \int_V \dd^3\bm{x}\,\partial_i\left(\Psi\partial^i \Psi_G-\Psi_G\partial^i  \Psi\right)\,. 
\end{align}
Using Green's theorem and calling ${\bf{S}}$ the oriented surface enclosing the volume $V$
\be\label{surf}
\Psi(0)=\int_S \dd{\bf{S}}\left(\Psi\partial^i \Psi_G-\Psi_G\partial^i\Psi\right) n_i\,,
\ee
where ${\bf{n}}$ gives the orientation at any point of the surface $S$ (with inward orientation). Note that this expression is valid for any surface $S$ with the characteristics listed above. The only assumption introduced to derive this is the thin lens approximation.

\section{Coefficients in Kirchhoff integral}\label{coeff}

We provide here the explicit expressions of the coefficients of the Kirchhoff integrand in the weak-field regime, entering the derivation of section~\ref{wf}.
\begin{align}
&\mathcal{K}_1= 8 \rho \Re(\gamma) -4 \rho \Re(\mu)- 2\rho^2 +2\Delta\rho -D\rho + 4 \Re(\delta \bar \tau) \nn \\
&\hspace{0,8cm} +4\tau\bar\tau -4\Re(\lambda\sigma)-2\sigma\bar\sigma +8\Re(\beta\bar\tau) \,, \label{B1}\\
&\mathcal{K}_2= 4\bar\sigma(\bar\beta-\bar\tau) -\bar\delta\bar\sigma \,, \label{B2}\\
&\mathcal{K}_3= 4\sigma(\beta-\tau) -\delta\sigma \,, \label{B3} \\
&\mathcal{K}_4= -4\sigma\bar\beta -4\rho\tau -2\delta\rho-\bar\delta\sigma \\
&\mathcal{K}_5= -4\bar\sigma\beta -4\rho\bar\tau -2\bar\delta\rho-\delta\bar\sigma\\
&\mathcal{K}_6= 16\rho^2\Re(\mu) +4\Re(\mu)D\rho -4\rho\Delta\rho -16 \rho^2\Re(\gamma) \nn \\
&\hspace{0,8cm} -16\rho\Re(\delta \bar \tau) -4\Re(\delta \bar \delta \rho) -2\Re(\delta \delta \bar \sigma +4\bar\sigma\delta\beta) \nn \\
&\hspace{0,8cm} +20\rho\Re(\lambda\sigma)+12\sigma\bar\sigma\Re(\mu) -32\rho\Re(\bar\beta\tau) +2\Re(\lambda D\sigma) \nn\\
& \hspace{0,8cm} -2\Re(\bar\sigma\Delta\sigma) -8\Re(\bar\beta\delta\rho) -4\Re(\bar\tau\delta\rho) -12\Re(\beta\delta\bar\sigma) \nn\\
& \hspace{0,8cm} +2\Re(\tau\delta\bar\sigma) -4\Re(\bar\sigma\delta\tau) \,,\label{K6}\\
&\mathcal{K}_7=\delta \bar \delta A_0-A_0 \rho \mu -\frac{1}{2}A_0\rho^2 \label{K7}\,.
\end{align}

\section{NP scalars solutions}\label{NPsol}

The solutions for the NP scalars are mostly derived in~\cite{Bruyere:2026gnt} and recalled here at linear order in the lens potential with the appropriate notations. 
\begin{align}
& \rho=-u+4\pi G\Sigma\left(\frac{u}{u_{\rm L}}\right)^2 \label{rhofirstorder}\\
& \mu=-\frac{u}{2}+2\pi G\Sigma\left[\left(\frac{u}{u_{\rm L}}\right)^2-\frac{u}{u_{\rm L}}\right]\\
& \delta\bar\tau=2\pi G\Sigma\frac{u^2}{u_{\rm L}}\\
& \delta\bar\delta\rho=2\pi G\nabla_\perp^2\Sigma\left(\frac{u}{u_{\rm L}}\right)^4-2\pi G\Sigma\frac{u^3}{u_{\rm L}}\left(1-2\frac{u}{u_{\rm L}}\right)\\
&\frac{\delta\bar\delta A_0}{A_0}=2\pi G\nabla_\perp^2\Sigma\frac{u^2}{u_{\rm L}^3}\left(1-\frac{u}{u_{\rm L}}\right) \\
&\hspace{1,2cm} +2\pi G\Sigma\frac{u^2}{u_{\rm L}}\left(1-2\frac{u}{u_{\rm L}}\right) \nn\\
& \frac{\Im\left(A_1\right)}{A_0}
= \pi G\nabla^2_{\perp}\Sigma\left(\frac{u_{\rm L}-u}{u_{\rm L}^2}\right)^2+2\pi G \Sigma  \left(\frac{u}{u_{\rm L}}\right)^2\label{A1A0u}
\end{align}
The expression for $\Im(A_1)$ is simply given by the results of Sec.~\ref{phasecorrection}.

Furthermore, we need the combination $\delta \delta \bar \sigma+4\bar\sigma\delta\beta$ to evaluate the coefficient \eqref{C2BE}. Since its evolution is governed by~\eqref{evddsig}, it can be written as
\begin{align}\label{ddsig}
\delta\delta\bar\sigma + 4\bar\sigma\delta\beta =\frac{1}{s^4}\int s^4 (\delta \delta \bar{\Psi}_0 +8 \delta \beta \bar \delta \bar \delta \phi)\dd s \,.
\end{align}
Therefore, we require the Weyl scalar $\delta \delta \bar{\Psi}_0$ at linear order, which is given by
\begin{align}\label{ddPsiob}
\delta \delta \bar{\Psi}_0 = &\delta \bar \delta \nabla^2 \phi -\frac{1}{2}\partial_{ss} \nabla^2 \phi-\frac{3}{s}\partial_s \nabla^2 \phi-\frac{2}{s^2}\nabla^2 \phi+\frac{1}{2}\partial_{ssss}\phi \nn\\
&+ \frac{4}{s}\partial_{sss}\phi + \frac{6}{s^2}\partial_{ss}\phi- 8 \delta \beta \bar \delta \bar \delta \phi \,.
\end{align}
Substituting this expression into the solution \eqref{ddsig}, integrating by parts, and using the Poisson equation, $\nabla^2\phi=4\pi G\rho_{\rm mat}$, one finds
\begin{align}
\delta\delta\bar\sigma + 4\bar\sigma\delta\beta = \frac{4 \pi G}{s^4} \int \left( s^4 \delta \bar \delta \rho_{\rm mat} + s^2\rho_{\rm mat} \right)\dd s \,.
\end{align}
In the thin lens approximation, we further use $\rho_{\rm mat}=\Sigma \,\delta(s-s_{\rm L})$ and $\delta \bar \delta \rho_{\rm mat}=\nabla^2_\perp\Sigma/2 \,\delta(s-s_{\rm L})$, which leads to
\be
\delta\delta\bar\sigma + 4\bar\sigma\delta\beta = 2 \pi G \nabla^2_\perp\Sigma \frac{u^4}{u_{\rm L}^4} + 4\pi G \Sigma \frac{u^4}{u_{\rm L}^2} \,.
\ee

\section{Point-like lens at second order}\label{PointLikeLens}

In vacuum, the first non-vanishing corrections to GO for a point-like lens appear at second order in the lens potential $\phi$, since the linear contribution in $GM$ vanishes. The dominant contribution was identified in \cite{Bruyere:2026gnt} as arising from $\delta\sigma$. Here, we recover this result directly from the structure of the Kirchhoff integral, providing an independent check of the equivalence between the two approaches.

The dominant scalars are those whose evolution equations contain the most efficient source terms. We identify them by examining their corresponding propagation equations. Since the first-order contribution vanishes in vacuum, we restrict our analysis to second order. We inspect the terms appearing in Eq.~\eqref{wave} and select the scalars that can generate the leading corrections.
Their sources are given by Weyl scalars, projected derivatives of Weyl scalars, or products of quantities that are themselves sourced by Weyl scalars. $\Psi_0$ is generically the dominant Weyl scalar as it does not involve any derivative along the affine parameter which suppress the contribution after integration along the geodesic.
Indeed, at linear order in the potential $\phi$, the Weyl scalars are given by
\begin{align}
\Psi_0=2\delta\delta\phi \,\,\,,\quad \Psi_1=-\partial_s\delta\phi \,\,\,,\quad \Psi_2=\frac{1}{2} \partial_{ss}\phi \,.
\end{align}
Consequently, among the NP scalars, $\sigma$ provides the leading contributions at first order since
\begin{align}
D\sigma= 2\rho\sigma +\Psi_0 \,.
\end{align}
In particular, in the thin lens approximation and at first order in the potential
\begin{align}\label{sigmafirstorder}
\sigma &= \left(\frac{u}{u_L}\right)^2 \int {\Psi_0} \dd s \nonumber\\
&=\frac{1}{2}\left(\frac{u}{u_L}\right)^2 (\partial_{xx} - \partial_{yy} + 2 \ii \partial_{xy})\psi\,.  
\end{align}
The same hierarchy applies to their projected derivatives.

Furthermore, the various second-order contributions do not scale in the same way. Some are of order $(GM/b)^2$, while others are enhanced by additional powers of $L/b$, which is larger than unity when the observer is located at a distance $L\gtrsim b$ from the lens. The dominant terms therefore scale as $(GM/b)^2 (L/b)^n$ with the largest possible integer $n$. Since all quantities entering the integrand are evaluated at the lens plane, where $L(s_{\rm L})=0$, the only factors of $L$ arise from the Gaussian integrals. At order $1/\omega$, the largest power is $L^3$, and it originates from the $(\zeta\bar\zeta)^3$ terms in Eq.~\eqref{wave}. The only second-order contribution entering this term is $\delta\sigma\bar\delta\bar\sigma$ arising from the product $\mathcal{K}_2\mathcal{K}_3$ [Eqs.~\eqref{B2},\eqref{B3}]. 
Proceeding as in \ref{linear}, we obtain the dominant correction from the Kirchhoff integral,
\begin{align}\label{A1A0final}
\omega\frac{\Im(A_1)}{A_0}^{(2)}\simeq\frac{1}{6}\delta\sigma^{(1)}\bar\delta\bar\sigma^{(1)}\Big\vert_{\rm L} L^3
\end{align}
Since $\delta\sigma$ enters quadratically, it is sufficient to compute it at first order. Keeping only the terms contributing to the dominant second-order correction, its evolution equation reduces to
\be
D\delta\sigma^{(1)} = 3\rho \delta\sigma + \delta\Psi_0 \,.
\ee
Defining ${\cal I}_{\delta\Psi_0} \equiv \int\delta\Psi_0\dd s$, and working in the thin lens approximation, we obtain
\begin{align}
\delta\sigma^{(1)}={\cal I}_{\delta\Psi_0}\left(\frac{s_{\rm L}}{s}\right)^3 \,.
\end{align}
Replacing in~\eqref{A1A0final}, we obtain the dominant second-order contribution in vacuum 
\be
\frac{\Im(A_1)}{A_0} \simeq \frac{1}{6}\left|{\cal I}_{\delta\Psi_0}\right|^2L^3\,.
\ee
As expected, we recover with the Kirchhoff integral the dominant solution obtained in section 3.4 of \cite{Bruyere:2026gnt} with a pure BE integration until the observer.

\bibliographystyle{unsrt}
\bibliography{biblio_Kirchhoff}

\end{document}